\documentclass[graybox]{svmult}

\usepackage{type1cm}        
\usepackage{makeidx}         
\usepackage{graphicx}        
\usepackage{multicol}        
\usepackage[bottom]{footmisc}
\usepackage{amsmath, amsfonts, amssymb, mathrsfs}

\usepackage{newtxtext}       %
\usepackage[varvw]{newtxmath}

\newcommand{\lp}{\left(}
\newcommand{\rp}{\right)}
\newcommand{\lb}{\left[}
\newcommand{\rb}{\right]}

\newcommand{\be}{\begin{equation}}
\newcommand{\ee}{\end{equation}}
\newcommand{\beqa}{\begin{eqnarray}}
\newcommand{\eeqa}{\end{eqnarray}}
\newcommand{\LL}{{\cal L}}

\renewcommand\l{\lambda}
\newcommand\m{\mu}
\newcommand\n{\nu}

\newcommand\s{\sigma}

\renewcommand\a{\alpha}
\newcommand\vk{\varkappa}
\newcommand\PP{{\cal P}}
\newcommand\OO{{\cal O}}
\newcommand\Det{{\rm Det\; }}
\def\D{\Delta}

\newcommand{\bseq}{\begin{subequations}}
\newcommand{\eseq}{\end{subequations}}

\usepackage{color}
\usepackage{braket}
\usepackage{mathrsfs}

\def\d{\partial}
\newcommand{\di}{\mathrm d}

\begin{document}

\title*{Ho\v{r}ava models as palladium of unitarity and renormalizability in quantum gravity}
\author{Andrei O. Barvinsky}
\date{}
\institute{Andrei O. Barvinsky \at Theory Department, Lebedev Physics Institute, Leninsky Prospect 53, Moscow 119991, Russia, and\\
Institute for Theoretical and Mathematical Physics, Moscow State University, Leninskie Gory, GSP-1, Moscow, 119991, Russia,
\email{barvin@td.lpi.ru}}
%
%
\maketitle


\abstract{We give a review of UV renormalization of Ho\v{r}ava gravity (HG) models introduced as a remedy against violation of unitarity in quantum gravity theory. Projectable and non-projectable low-dimensional HG models and the spectra of their physical degrees of freedom are considered in the linearized approximation on flat spacetime background. The problem of regularity of their propagators depending on the choice of gauge fixing procedure is discussed along with the role of this regularity in UV renormalization. With the choice of a special class of quasi-relativistic gauge conditions perturbative renormalizability of projectable HG models is proven in any spacetime dimension and the status of renormalization in non-projectable models is briefly discussed. We show how the covariance of counterterms is provided within the class of background covariant gauge conditions and show asymptotic freedom of $(2+1)$-dimensional HG. We also present the calculation of beta-functions in $(3+1)$-dimensional theory by the generalized Schwinger-DeWitt technique of universal functional traces and obtain fixed points of its renormalization group flow, some of them being good candidates for asymptotic freedom.}


\section{Introduction}
\label{sec:1}
The quest for renormalizable and perturbatively consistent in UV domain quantum gravity theory shows that such a theory can indeed be constructed by introducing higher order curvature invariants \cite{Sch-DeWitt,Stelle:1976gc,Fradkin:1981hx,Avramidi:1985ki}, but it is doomed to violate unitarity in view of ghost modes associated with higher order derivatives. In spite of various efforts to circumvent this problem or justify the presence of ghosts by special rules of handling them \cite{Salvio:2014soa,Einhorn:2014gfa,Anselmi-Piva,Mannheim} or within the scope of string theory, nonlocal field models \cite{KKStar,LucaB}, etc., the most widespread point of view is that absence of ghosts should be a criterion of selecting a healthy stable theory, and if this theory is also local, renormalizable, unitary and consistent in UV limit, then there is a hope that it can also describe our Nature.

Here we discuss the mechanism that can provide a combination of these properties, based on the breakthrough suggestion of \cite{Horava:2008ih,Horava:2009uw} that this can be achieved by dropping the requirement of Lorentz invariance and introducing in the field theory higher order derivatives only with respect to spatial coordinates. This suggestion turned out to be very productive and used the notion, borrowed from condensed matter physics \cite{Lifshitz}, of Lifshitz anisotropic scaling and scaling dimensions. Anisotropic scaling dimension replaces conventional physical dimensionality as a criterion of convergence of Feynman diagrams. Application of this criterion within simple power counting arguments leads in \cite{Horava:2009uw} to an invention of the class of local, unitary quantum gravity models which are considered to be perturbatively renormalizable and, therefore, expected to be consistent in UV domain.

The anticipation of such a fundamental breakthrough in high-energy domain of quantum gravity served as a motivation to explore low-energy
consistency and phenomenology of Ho\v rava's proposal, see
\cite{Mukohyama:2010xz,Sotiriou:2010wn} for reviews. In particular, this has led to the identification of a Ho\v{r}ava gravity (HG) version  ---
the so-called healthy non-projectable model~\cite{Blas:2009qj} --- which provides a consistent theory reproducing the phenomenology of general
relativity (GR) at the distance scales where the latter has been tested. It was also realized that the theory never reduces to GR exactly: a certain amount of Lorentz invariance violation persists in the gravity sector at all distance scales \cite{Blas:2010hb}. This might have interesting implications for cosmological models of dark energy \cite{Blas:2011en} and lead to astrophysical and cosmological constraints on the parameters of the theory
\cite{Audren:2014hza,Yagi:2013qpa,Blas:2014aca}. To  be phenomenologically viable, this model should include a mechanism ensuring Lorentz invariance in the sector of visible matter --- the challenging issue addressed in \cite{GrootNibbelink:2004za,Pujolas:2011sk,Pospelov:2010mp,Bednik:2013nxa,
Kharuk:2015wga}.

Despite vast literature on HG there still remained a number of subtle issues in the fundamentals of Ho\v{r}ava proposal \cite{Horava:2009uw}. Namely, this are the problem of irregular propagators which might break the conventional BPHZ scheme of subtracting UV divergences by local counterterms and the problem of local gauge invariance of these counterterms, both of these issues especially inherent in HG construction. Point is that a general local gauge fixing in HG induces certain ``irregular'' contributions in the propagator of the metric, that may spoil the convergence of the loop integrals \cite{Anselmi:2007ri}. As a consequence, a loop diagram that by a scaling argument should be finite can actually diverge and generate a counterterm not expected from the naive power-counting or, even, give rise to a nonlocal divergence. The key question was whether there exists a class of gauges where all propagators are regular.

Another problem is the issue of covariant counterterms. It is well known that their manifest covariance as well as manifestly covariant nature of all intermediate calculations can be enforced by the use of the background field formalism in the class of background covariant gauges. For various field models this was demonstrated within one-loop approximation \cite{DeWitt_covariant,Veltman:1975vx,Abbott:1981ke}, generically in perturbation theory \cite{Voronov:1,Voronov:2} and in the framework of BRST cohomology methods with a special account of locality properties \cite{Barnich:1994ve,Barnich:1994mt,Barnich:1995ap}. However, the extension of these results to local gauge theories with broken Lorentz invariance and effective field theories was not yet known. Clear demonstration that the BRST structure of renormalization in HG and other models with similar gauge invariance algebra is such that it reduces to renormalization of physical gauge fields separately from the renormalization of the BRST ghost sector was missing.

Both of the above problems have been successfully solved for the class of {\em projectable} Ho\v rava models which have been proven to be perturbatively renormalizable in any spacetime dimension \cite{HG,BRST}. Moreover, in the series of papers \cite{2+1,towards,3+1} $(2+1)$-dimensional projectable HG was shown to be asymptotically free in UV limit, while its $(3+1)$-dimensional version turned out to have several interesting fixed points of renormalization group (RG) flow that can also be good candidates for asymptotic freedom. The goal of this paper is to give a brief review of these results.

Unfortunately, the projectable version of Ho\v rava gravity does not
reproduce GR at low energies (at least not within weak
coupling)~\cite{Blas:2010hb}. Nevertheless, it presents an interesting example of a theory sharing many properties of GR, such as a large gauge group of local spacetime transformations and the presence of gapless
transverse-traceless excitations --- gravitons --- in dimensions $d=3$
and higher. Working in the gauge with regular propagators we demonstrate
that projectable Ho\v rava gravity is perturbatively renormalizable in the strict sense.

For non-projectable HG model, which both theoretically and phenomenologically can be consistent with GR in the infrared domain \cite{Blas:2010hb}, the regularity conditions of its propagators turn out to be violated. But these conditions are only sufficient rather than necessary for renormalizability, because a subtle mechanism of cancellation of harmful irregularities remains possible, as it was recently shown in \cite{Belorin2022}. So we also briefly discuss the source of this problem originating from the peculiarities of the canonical formalism of non-projectable HG.

Renormalization of HG models undertaken in \cite{2+1,towards,3+1} raises another problem -- enormous computational complexity associated with humongous amount of Feynman diagram vertices caused by the lack of Lorentz and full $(d+1)$-dimensional diffeomorphism invariance of the theory. While in the renormalization of $(2+1)$-dimensional HG it was possible to use standard Feynman diagrams to reach the result \cite{2+1}, in a similar $(3+1)$-dimensional case this becomes virtually impossible. It is enough to say that the inverse metric propagator in the background field formalism amounts to several hundred terms. So the method based on the combination of background field formalism and heat kernel technique \cite{Schwinger,Sch-DeWitt,DeWitt:2003pm,PhysRep,twoloop,Scholarpedia} becomes indispensable. This method provides the UV divergences not as expansion in powers of field perturbations, but as full nonlinear counterterms --- local nonlinear functionals of the generic background field. Pioneering application of this method in Einstein theory \cite{tHooft-Veltman} proved to be very efficient and now underlies the majority of results on renormalization of
(super)gravitational models. The basic tool of this method is
the heat equation kernel whose proper time expansion coefficients ---
the so-called HAMIDEW \cite{Gibbons} or Gilkey--Seeley coefficients ---
carry a full information about UV divergences and can be
systematically calculated.

Despite powerful calculational advantages of the heat kernel method,
its application to HG encounters the following major difficulty.
It is directly applicable to the so-called {\em minimal} operators --- second order differential operators in which all spacetime derivatives are treated on equal footing and form covariant d'Alembertians. Existence of preferred time foliation in HG obviously violates this property. Several approaches have been put forward to circumvent this problem and extend the heat kernel method to Lifshitz-type theories
\cite{Nesterov:2010yi,DOdorico:2014tyh,DOdorico:2015pil,HKLT, Grosvenor:2021zvq,Saueressig}.
Another difficulty in applications to HG models --- {\em non-minimal} operators arising in these models have higher order derivative terms which are also not exhausted by powers of the spatial Laplacian
$\Delta\equiv\gamma^{ij}\nabla_i\nabla_j$. The principal symbol term of
these operators is non-diagonal in derivatives whose indices are contracted with the tensor field indices. This difficulty was circumvented in \cite{3+1} by the generalized Schwinger-DeWitt technique of the so-called {\em universal functional traces} (UFT), that was originally developed for spacetime covariant operators in \cite{PhysRep,twoloop} (see also \cite{JackOsborn}). Here we will give a brief overview of this technique used for the calculation of the full set of beta-functions in $(3+1)$-dimensional projectable HG.

The paper is organized as follows. We begin with the Lifshitz idea of scaling which is anisotropic between space and time \cite{Lifshitz} and then apply it in Sect.\ref{HG-models} to quantum gravity as a remedy against violation of unitarity in the form of HG theory. After formulating the foliation preserving diffeomorphism symmetry of HG models we consider them in the linearized approximation on flat spacetime background along with their spectra of physical degrees of freedom in two low-dimensional cases. Then in Sect.\ref{reg_prop} we discuss the problem of regularity of their propagators depending on the choice of gauge fixing procedure and the role of this regularity in UV renormalization. With the choice of a special class of quasi-relativistic gauge conditions we prove perturbative renormalizability of projectable HG models in any spacetime dimension and briefly dwell on the status of renormalization in non-projectable models. In Sect.\ref{BRST} we show how covariance of counterterms is provided within the class of background covariant gauge conditions and then go over to the calculation of one-loop counterterms and renormalization group beta-functions in two low-dimensional theories. In this way in Sect.\ref{AF2+1} we show asymptotic freedom of $(2+1)$-dimensional HG and present in Sect.\ref{3+1} the calculation of beta-functions for $(3+1)$-dimensional theory by the generalized Schwinger-DeWitt technique of universal functional traces. After a brief discussion of fixed points and their properties we finish the paper with a concluding discussion.

\section{Lifshitz theories with anisotropic scaling}
As is well known, quantum gravity can be rendered UV renormalizable via introducing curvature-squared counterterms due to simple power-counting arguments of DeWitt \cite{Sch-DeWitt,DeWitt_covariant} later justified by a rigorous analysis of \cite{Stelle:1976gc}. This however leads to higher order spacetime derivatives in the Lagrangian and inevitably leads to ghost instabilities and loss of unitarity. The idea of salvation of unitarity in quantum gravity \cite{Horava:2008ih,Horava:2009uw} comes from Lifshitz work on phase transitions in condensed matter physics \cite{Lifshitz} suggesting the anisotropy between space and time. The ghost modes violating unitarity originate entirely from higher-order time derivatives. On the other hand, the UV convergence of Feynman diagrams can be improved by introducing higher order derivatives only with respect to spatial coordinates, thus making the originally nonrenormalizable theory renormalizable without violation of unitarity.

The implementation of this idea can be demonstrated on the example of Lifshitz scalar theory with the anisotropy between time and space. Consider the transition from usual relativistic invariant action to the action of a scalar field in $(d+1)$ spacetime dimensions
\begin{equation}
    S=-\int \di^{d+1}x\,\partial_\mu\phi\,\partial^\mu\phi\,\,\Longrightarrow S_L=\int
     \di t\,\di^dx(\dot\phi^2+\phi\,{\bf D}\phi),
\end{equation}
where ${\bf D}$ is a higher order differential operator in spatial derivatives of the form
\begin{equation}
    {\bf D}=-\frac{(-\Delta)^z}{(M^2)^{z-1}}+\dots, \quad \Delta=\d_i\d^i,\quad z>1.
\end{equation}
Here the mass parameter $M^2$ is introduced to keep the {\em physical} dimensionality correct and lower derivative terms are denoted by dots.

In contrast to Lorentz symmetry and usual scaling invariance the new action
becomes invariant under the symmetry broken down to $O(d)$ and a special {\em anisotropic} scaling,
\begin{equation}\label{scaling}
    \begin{cases}
    x^{\mu}\mapsto b^{-1}x^{\mu}\\
    \phi\mapsto b^{\frac{d-1}{2}}\phi,~~
    [\,\phi\,]=\frac{d-1}{2}
    \end{cases}
    \!\!\!\!\Longrightarrow\,\,\,\,
    \begin{cases}
    t\mapsto b^{-z}t,~~~
    x^i\mapsto b^{-1}x^i,\\
    \phi\mapsto b^{\frac{d-z}{2}}\phi,~~
    [\,\phi\,]=\frac{d-z}{2},
    \end{cases}
\end{equation}
which allows one to introduce the notion of {\em anisotropic} scaling dimension of the field --- a field $\phi$ with dimension $[\phi]=r$ transforms under the new scaling (\ref{scaling}) as $\phi\mapsto b^{r}\phi$. Accordingly we assign dimension $-1$ to the spatial coordinates $x^i$ and time has dimension $-z$. Note that for $z\neq 1$ a scaling dimension is \textit{no longer} equal to a physical dimension.

As will be clear in what follows, at the quantum level the renormalizability of such theories is determined by the same dimensional power counting arguments as in Lorentz invariant theories, but the role of dimensionality should be played by the scaling dimensionality rather than the physical one. This allows one to formulate a simple criterion for the choice of the parameter $z$ for which the theory becomes renormalizable.

It is natural that for nonlinear self-interacting fields the interaction terms contain highest spatial derivatives of the same order $2z$, say in the cubic order in $\phi$ of a type $\sim\lambda\partial^{2z}\phi^3$. Therefore, in view of the dimension of the integration measure $[\di t\,\di^dx]=-z-d$ and zero dimension of the interaction action, which we assume because the scaling symmetry is not supposed to be broken,
\begin{equation}
    [\lambda\,\partial^{2z}\phi^3]=[\lambda]+2z+\frac{3}{2}(d-z)=z+d,
\end{equation}
the requirement of renormalizability --- non-negative dimension of the coupling constant $\lambda$ --- reads
\begin{equation}
    [\lambda]=\frac{z-d}{2}\geq 0.
\end{equation}
This leads to the critical value of $z$ at which the theory becomes renormalizable (and superrenormalizable beyond this value), $z_{\mathrm{crit}}=d$.

Let us now apply this idea in quantum gravity theory.

\section{Ho\v{r}ava gravity models \label{HG-models}}

The idea of Lifshitz anisotropic scaling implies an obvious mismatch between the number of spatial and temporal derivatives and leads to the loss of Lorentz invariance. In case of gravity this means that the theory cannot retain full local diffeomorphism invariance, and this local symmetry should be chosen to respect this higher derivative structure in space vs two derivatives in time. Such a symmetry can obviously be associated with the ADM split of the gravitational configuration space into spatial metric $\gamma_{ij}$, lapse $N$ and shift $N^i$ functions,
 \begin{eqnarray*}
ds^2=-N^2\di t^2+\gamma_{ij}(\di x^i+N^i\di t)(\di x^j+N^j\di t)~,~~i,j=1,\ldots,d,
\end{eqnarray*}
where $d$ is the dimensionality of space, which we consider rather general in order to learn how the properties of the model depend on spacetime dimensionality $D=d+1$. Under this split the minimal truncation of the diffeomorphism invariance looks like the so-called foliation preserving diffeomorphisms FDiffs under which spatial coordinates undergo generic time dependent transformations accompanied by space independent reparametrization of time,
    \begin{equation}
    t\mapsto\tilde t=\tilde t(t),\quad
    x^i\mapsto\tilde x^i=\tilde x^i(\textbf{x},t), \label{x_to_tilde_x}
    \end{equation}
Lapse function, shift functions, spatial metric and the extrinsic curvature
    \begin{equation}
    K_{ij}=\frac{1}{2N}\lp\dot\gamma_{ij}-\nabla_iN_j-\nabla_jN_i\rp
    \end{equation}
transform in such a way that the transformation laws for $\gamma_{ij}$ and $K_{ij}$ have a homogeneous tensor type nature,
\begin{eqnarray}
\label{Ftrans}
&&N\mapsto \tilde N=N\frac{\di t}{\di\tilde t}~,\quad
N^i\mapsto \tilde N^i=\bigg(N^j\frac{\d\tilde x^i}{\d x^j}
-\frac{\d\tilde x^i}{\d t}\bigg)\frac{\di t}{\di\tilde t}~,\\
\label{Ftrans1}
&&\gamma_{ij}\mapsto\tilde\gamma_{ij}=\gamma_{kl}
\frac{\d x^k}{\d\tilde x^i} \frac{\d x^l}{\d\tilde x^j}~,\quad
K_{ij}\mapsto\tilde K_{ij}=K_{kl}
\frac{\d x^k}{\d\tilde x^i} \frac{\d x^l}{\d\tilde x^j}~.
\end{eqnarray}

This easily enables to construct the invariant kinetic term of the action at most quadratic in time derivatives as a generic quadratic form in $K_{ij}$ and the rest of its Lagrangian can be constructed from local operators that transform as scalars under FDiffs and have dimension up to $2d$ -- the maximal order of spatial derivatives,
    \begin{equation}
    S_{\mathrm{HG}}[\gamma_{ij},N^j,N]=\frac1{2G}\int \di t\,\di^dx\,N\,\gamma^{1/2}\big(K^2_{ij}
    -\lambda K^2-\mathcal{V}\big).           \label{genact}
    \end{equation}
Here $\lambda$ and $1/G=M^{d-1}$ are coupling constants (the latter for (3+1)-dimensional case is associated with the Planck mass squared $M_{\rm P}^2$), $K=\gamma^{ij}K_{ij}$, the dot stands for a time-derivative, indices are raised and lowered by the spatial metric
$\gamma_{ij}$ and the covariant spatial derivatives $\nabla_{i}$ are compatible with $\gamma_{ij}$. The potential term ${\cal V}$ consists of all allowed combinations of local invariants of scaling dimension up to $2d$ that are made of $\gamma_{ij}$, $N$ and their covariant derivatives $\nabla_{i}$. In this way one has a Lagrangian consisting of marginal and relevant operators with respect to the anisotropic scaling which in this sense is at least naively power-counting renormalizable if one prescribes the following scaling dimensions to the full set of field variables,
    \begin{equation}
    [\gamma_{ij}]=[N]=0,\quad [N^i]=d-1,\quad [K_{ij}]=d,\quad[\mathcal{V}]=2d.
    \end{equation}

With this choice, corresponding to the value $z=d$ of the parameter $z$ introduced above, the action has zero scaling dimension, $[S_{\mathrm{HG}}]=0$, in view of $[\di ^dx]=[\di t]=-d$. Note that both coupling constants $G$ and $\lambda$ also have zero {\em scaling} dimension in contrast to the situation with the {\em physical} dimension of $G$, $[\,G\,]=0$ vs $[\,G\,]_{\rm phys}=1-d$.

In the {\it non-projectable} Ho\v rava gravity the lapse $N$ is postulated
to be a function of both space and time. The status of renormalizability of this model is special, so that we postpone the discussion of
this case until Sec.~\ref{Non-projectable}. So, to begin with, we focus on the {\it projectable} model where the lapse is a function of time only,
$N=N(t)$. Then the time reparameterization invariance allows one to set $N=1$ leaving the time-dependent spatial diffeomorphisms as the remaining gauge
transformations.

\subsection{Projectable models}
For projectable models the potential term of the Ho\v{r}ava Lagrangian is a local function of spatial metric, curvature tensor and its covariant derivatives. In the $(2+1)$-dimensional case, $d=2$, the potential includes only two terms,
\be
\label{Vd2}
{\cal V}_{d=2}=2\Lambda+\m R^2\;
\ee
because the linear in $R$ term is a total derivative in 2-dimensions and the
Ricci tensor $R_{ij}=\tfrac12\gamma_{ij}R$ reduces to the scalar curvature. Setting the cosmological constant $\Lambda$ to zero, we obtain a model with three marginal couplings $G$, $\l$ and $\m$.

The second term of (\ref{Vd2}) together with the extrinsic-curvature terms
are marginal under the scaling (\ref{scaling}). They determine the UV
behavior of the theory, in particular its renormalizability
properties. The cosmological constant term with $\Lambda$ is a relevant
deformation of the lowest dimension, which breaks anisotropic scaling in the infrared limit. We assume that it is tuned to zero in order to admit flat Minkowski spacetime as a solution.

To study the spectrum of linear perturbations around this
background we write
\be
\label{metrpert}
\gamma_{ij}=\delta_{ij}+h_{ij},
\ee
and decompose the perturbations into scalar, vector and transverse-traceless (TT) tensor parts,
\begin{eqnarray}
    &&h_{ij}=\lp\delta_{ij}-\frac{\partial_i\partial_j}{\Delta}\rp\psi
    +\frac{\partial_i\partial_j}{\Delta}E
    +2\partial_{(i}v_{j)}+t_{ij},\quad
    N^i=\partial^i B+u^i,    \label{metrlin}\\
    &&\partial^iu_i=\partial^iv_i=0,
    \quad t^i_i=0=\partial^jt^i_j.
\end{eqnarray}

Expanding around flat spacetime and performing this
decomposition we obtain the quadratic action,
\beqa
\label{S2d2}
S^{(2)}_{d=2}=\frac{1}{2G}\int \di t\,\di ^2x
&\bigg[&-\frac{1}{2}(\dot v_i-u_i)\D(\dot v_i-u_i)
+\frac{\dot\psi^2}{4}+\frac{1}{4}(\dot E-2\D B)^2\nonumber\\
&&-\frac{\l}{4}(\dot\psi+\dot E-2\D B)^2-\m \psi\D^2\psi\,\bigg]\;.
\eeqa
Variation with respect to $u_i$ implies that there are no propagating
modes in the vector sector. In the scalar sector we eliminate $E$ using the
equation obtained upon variation with respect to $B$ and set as a gauge condition $B=0$ afterwards. This yields
\be
\label{S2d2phys}
S^{(2)}_{d=2}=\frac{1}{2G}\int
\di t\,\di ^2x\,\bigg[\,\frac14\,\frac{1-2\l}{1-\l}\dot\psi^2
-\m \psi\D^2\psi\,\bigg]\;,
\ee
so that unlike GR, which in $(2+1)$ dimensions does not possess any local
degrees of freedom, Ho\v rava gravity propagates a dynamical scalar
mode. The latter has the dispersion relation,
\be
\label{dispreld2}
\omega^2_s=4\mu \frac{1-\l}{1-2\l}\,k^4\;.
\ee
It is well-behaved (i.e. has positive kinetic term and is stable)
if $G>0$, $\m>0$ and $\l<1/2$ or $\l>1$.

In $d=3$, upon using the Bianchi identities, integrating by parts and noting that the Riemann tensor expresses in terms of the Ricci one,
one finds the most general potential~\cite{Sotiriou:2009gy},
\be
\label{Vd3}
\begin{split}
{\cal V}_{d=3}=&\;2\Lambda-\eta R+\m_1R^2+\m_2R_{ij}R^{ij}\\
&+\n_1R^3+\n_2RR_{ij}R^{ij}+\n_3R^i_jR^j_kR^k_i
+\n_4\nabla_iR\nabla^iR+\n_5\nabla_iR_{jk}\nabla^iR^{jk}\;.
\end{split}
\ee
Here, $R_{ij}$ and $R$ are the Ricci tensor and Ricci scalar
constructed from $\gamma_{ij}$.
In total, the theory contains 11 couplings: $G$, $\l$, $\Lambda$,
$\eta$, $\m_{1,2}$ and $\n_a$, $a=1,\ldots,5$. The terms in the second
line of (\ref{Vd3}) together with the extrinsic-curvature terms in (\ref{genact})
are marginal under the scaling (\ref{scaling}). They determine the UV
behavior of the theory, in particular its renormalizability
properties. The rest of the terms in (\ref{Vd3}) are relevant
deformations. Among them the cosmological constant $\Lambda$, which has the
lowest dimension.

Setting $\varLambda$ again to zero and expanding the action around flat Minkowski spacetime we get its quadratic part
\begin{equation}
    \begin{split}
        S^{(2)}_{d=3}=\frac1{2G}\int \di t\,\di^3x\,&\bigg\{\,\bigg[\, \frac{\dot t_{ij}^2}{4}+\frac{\eta}{4}t_{ij}\Delta t_{ij}-\frac{\mu_2}{4}t_{ij}\Delta^2t_{ij}+\frac{\nu_5}{4}t_{ij}\Delta^3t_{ij} \bigg]\\
        &-\frac{1}{2}(\dot v_i-u_i)\Delta(\dot v_i-u_i)\\
        &+\frac{\dot\psi^2}{2}+\frac{1}{4}(\dot E-2\Delta B)^2
        -\frac{\lambda}{4}(2\dot\psi+\dot E-2\Delta B)^2\\
        &-\frac{\eta}{2}\psi\Delta\psi-\Big(4\mu_1
        +\frac{3\mu_2}{2}\Big)\,\psi\Delta^2\psi
        +\Big(4\nu_4+\frac{3\nu_5}{2}\Big)\,
        \psi\Delta^3\psi\bigg\},                    \label{S2d3}
    \end{split}
\end{equation}
where the first line represents the tensor sector of transverse traceless gravitons, the second sector corresponds to vector modes and the last two lines form the scalar sector.

In order to identify the physical degrees of freedom we perform the variation
with respect to $u_i$ and $B$ and set them to zero afterwards by the gauge
choice (three gauge conditions $u_i=B=0$ for three diffeomorphisms). We obtain the equations,
\be
\label{uBeqd3}
\D\dot v_i=0~,~~~~\D\Big(\dot E-\frac{2\l}{1-\l}\dot\psi\Big)=0\;.
\ee
The first one implies that the vector sector again does not contain any
propagating modes. From the second equation in (\ref{uBeqd3}) we
express $\dot E$ and substitute it back into (\ref{S2d3}) which yields
the action for the propagating modes,
\begin{equation}
    \begin{split}
        S^{(2)}_{d=3}&=\frac{M_P^2}{2}\int \di t\,\di^3x\,\bigg\{\,\frac{\dot t_{ij}^2}{4}+\frac{t_{ij}}{4}
        \lb\, \eta\Delta-\mu_2\Delta^2+\nu_5\Delta^3\rb t_{ij}\\
        &+\frac{1-3\lambda}{1-\lambda}
        \frac{\dot\psi^2}{2}+\frac{1}{2}\psi\lb -\eta\Delta-(8\mu_1+3\mu_2)\Delta^2
        +(8\nu_4+3\nu_5)\Delta^3\rb\psi\bigg\}~.   \label{S2d3phys}
    \end{split}
\end{equation}

In addition to the TT mode $t_{ij}$, the theory propagates a ``scalar
graviton'' $\psi$. Both modes have positive-definite kinetic terms
provided $G>0$ and $\l$ is either smaller than $1/3$ or larger
than $1$. The dispersion relations of two transverse-traceless modes and one scalar mode $\propto e^{-i\omega t+i\textbf{k}\textbf{x}}$ are respectively,
\beqa
\label{dispreld3tt}
&&\omega_{tt}^2=\eta k^2+\m_2k^4+\n_5k^6\;,\\
\label{dispreld3s}
&&\omega_{s}^2=\frac{1-\l}{1-3\l}\big(-\eta k^2+(8\m_1+3\m_2)k^4
+(8\n_4+3\n_5)k^6\big)\;.
\eeqa
This immediately raises a problem: the $k^2$-term in the dispersion relation cannot be positive for both modes simultaneously. Thus,
non-zero $\eta$ leads to an instability of the Minkowski background
with respect to inhomogeneous perturbations. For positive values of the
parameters $\m_{1,2}$ and $\n_{4,5}$ the instability is cut off at
large spatial momenta and therefore does not affect the UV properties
of the theory. Moreover, we can stabilize the Minkowski spacetime by
simply tuning $\eta$ to zero. However, in that case the dispersion
relations of the TT mode and scalar gravitons are quadratic,
$\omega\propto k^2$, down to zero momentum, which prevents from
recovering GR at low energies\footnote{One could try suppressing the instability with finite and positive $\eta$ by tuning $\l$ close to $1$. However, in this limit the theory becomes strongly coupled and the perturbative treatment breaks down~\cite{Blas:2009yd,Blas:2010hb}.}.

We will be waving aside this difficulty of matching in the low energy domain the dispersion relations with those of general relativity \cite{Blas:2010hb}, because we will restrict ourselves only with the analysis of renormalizability of the theory in high-energy domain. This analysis usually proceeds in the ``Euclidean'' time obtained by the Wick rotation $t\mapsto\tau=it$, $N^j\mapsto N^j_{\rm E}=-iN^j$,  and the use of the relation $iS=-S_{\rm E}$ between the initial action and the Euclidean action $S_{\rm E}$. The corresponding Euclidean action then differs from (\ref{genact}) only by the replacement of $t$ by $\tau$ and the sign of the potential term. At the quadratic level this amounts to flipping the signs of the terms containing $\m_{1,2}$, $\n_{4,5}$ in (\ref{S2d3}) and of the $\m$-term in (\ref{S2d2}).

\section{Renormalizability and the problem of irregular propagators \label{reg_prop}}

The prospects of renormalization of the above models turn out, however, more complicated than it was originally anticipated in \cite{Horava:2009uw}, because the proof of renormalizability cannot really be accomplished by naive power counting arguments. Point is that the degree of divergence of Feynman diagrams (denoted by $\cal D_{\rm div}$) does not a priori provide correct renormalizability criteria of the BPHZ mechanism, because in the transition from Lorentz invariant theories to Ho\v rava models it is now based on counting the anisotropic scaling dimension of their typical integrands,
\begin{equation}
{\cal D}_{\rm div}\left(\int\frac{d^{d+1}p}{(p^2)^N}\right)=1+d-2N\Rightarrow {\cal D}_{\rm div}\left(\int\frac{d\omega\,d^{d}{\bf k}}{(A\omega^2+B{\bf k}^{2z})^N}\right)=z+d-2zN,
\end{equation}
with rather general coefficients $A$ and $B$. Some of these coefficients might be zero and this creates a serious problem.

More generally this transition to the Lorentz non-invariant integrals over $(d+1)$-dimensional loop momenta $p=(\omega,{\bf k})$,
\begin{align}
&\int \prod_{l=1}^L \di^{d+1}p^{(l)}\;
{\cal F}_n(p)\;
\prod_{m=1}^M \frac1{\big(P^{(m)}(p)\big)^2}\nonumber\\
&\quad\Rightarrow
\int \prod_{l=1}^L \di\omega^{(l)}\di^d{\bf k}^{(l)}\;
{\cal F}_n(\omega,{\bf k})\;
\prod_{m=1}^M \frac1{{A_m}\big(\Omega^{(m)}(\omega)\big)^2
+{B_m}\big({\bf K}^{(m)}({\bf k})\big)^{2z}},             \label{reg_int}
\end{align}
results in propagators with various constant coefficients $A_m$ and $B_m$ (here $\Omega^{(m)}(\omega)$ and ${\bf K}^{(m)}({\bf k})$ represent the momenta flowing in propagators as linear combinations of the full set of independent loop momenta $(\{\omega\},\{{\bf k}\})$). It turns out that the rules of BPHZ subtraction of UV divergences are guaranteed only when $A_m$ and $B_m$ are both positive \cite{HG}.

To clarify the origin of this difficulty first note that a generic diagram contains subdivergences and thus can diverge despite ${\cal D}_{\rm div}<0$. Fortunately, as shown in \cite{Anselmi:2007ri}, the combinatorics of the subtraction procedure in non-relativistic theories works essentially in the same way as in the relativistic case, and subdivergences are subtracted by
counterterms introduced at the previous orders of the loop expansion.
However, even in the absence of subdivergences, the convergence of a
diagram with ${\cal D}_{\rm div}<0$ is not trivial. Indeed, consider the $L$-loop integral
\be
\int \di \omega\,\di^d k\int \Big[\prod_{l=2}^L
\di \omega^{(l)}\di^d k^{(l)}\Big] \;
f(\{\omega\},\{{\bf k}\})\equiv
\int \di \omega\,\di^d k\,\tilde f\big(\omega,k\big)
\;,
\ee
where we singled out from the full set $(\{\omega\},\{{\bf k}\})$ the first loop momentum $(\omega,{\bf k})$ and suppressed the dependence on external momenta. Assume for simplicity that $f$ is a scalar function (in general it can carry tensor
indices corresponding to the external legs of the diagram). Because subdivergences are absent, the inner integral converges and gives a function
$\tilde f\big(\omega,k\big)$ which for ${\bf k}\mapsto b\,{\bf k}$, $\omega\mapsto b^2\,\omega$ scales as $b^{D_{\rm div}-2d}$.
However, the latter can have the form
\be
\label{spuriousdiv}
\tilde f(\omega,k) \sim \omega^{-1\pm n}
k^{{\cal D}_{\rm div}-d\mp dn}
~,~~~~n>0\;,
\ee
and the integral over frequency (momentum) will diverge, despite the
fact that the $k$-integral ($\omega$-integral) is finite. These are precisely the spurious divergences that arise if the propagators contain irregular contributions. Note that this problem is absent in Lorentz invariant theories, where the function $\tilde f$ can depend only on $\omega^2+k^2$. In \cite{HG} it was proven that spurious divergences of the form (\ref{spuriousdiv}) do not appear if all propagators in (\ref{reg_int}) have the regular form with all $A_m,B_m>0$. In that case ${\cal D}_{\rm div}<0$ indeed implies convergence of the diagram.\footnote{The exact statement of \cite{HG} is as follows. Consider a diagram with $L$ loops and ${\cal D}_{\rm div}<0$. Assume that all propagators in the diagram are regular in the sense (\ref{reg_int}) and that if the momentum and frequency in any of the propagators are frozen then the integral over remaining momenta and frequencies converges (i.e. subdivergences are absent). Then the whole diagram converges. This is the statement about convergence in the UV. Infrared divergences present a separate issue and must be regulated by an IR cutoff.}

The values of $A_m$ and $B_m$, however, depend on the choice of the model and, moreover, on the choice of gauge conditions used in the Faddeev-Popov (or BRST) gauge fixing procedure. This can be easily shown, say for the $d=2$ case, by inverting the Hessian of the action (\ref{S2d2}) in the degenerate gauge $N^i=0$. The $(ij,kl)$-block of the full propagator in the momentum representation, $p=(\omega,{\bf k})$, then contains the term
\be
\langle
h_{ij}(p)h_{kl}(-p)\rangle=
\big(\delta_{ik}\delta_{jl}+\delta_{il}\delta_{jk}
-2\delta_{ij}\delta_{kl}\big)\frac{2G}{\omega^2}+...,
\ee
corresponding in the coordinate representation to the kernel $\langle h_{ij}(\tau,{\bf x})h_{kl}(\tau',{\bf x'})\rangle=
-G\big(\delta_{ik}\delta_{jl}+\delta_{il}\delta_{jk}
-2\delta_{ij}\delta_{kl}\big)\, |\tau-\tau'|\,\delta^{(2)}({\bf x}-{\bf x'})+\dots$, which is singular for all $\tau-\tau'$. This is in sharp contrast to local point-like singularities in Euclidean spacetime at $x=x'$ providing renormalization of UV divergences by local counterterms.

Another example is the nondegenerate gauge corresponding to the addition to the gauge invariant Lagrangian of the gauge-breaking term
\be
\label{local}
\LL_{\rm gf}=\frac{\sigma}{2G}\,F^i{\cal O}_{ij}F^j\;,
\ee
where $F^i$ is a set of gauge condition functions linear in fields $N^i$, $h_{ij}$ and their derivatives, while ${\cal O}_{ij}$ is an invertible gauge-fixing operator and $\sigma$ is a relevant gauge-fixing parameter. In order not to spoil the
scaling properties of the action, this gauge-fixing term should have the total dimension of $2d$, whereas all terms in $F^i$ and ${\cal O}_{ij}$ must scale appropriately. For the example of $d=2$ model with the nondegenerate gauge on shift vector variables, $F^i=N^i$, the obvious choice is $\OO_{ij}=-\delta_{ij}\D-\xi\d_i\d_j$, and this leads to the propagator block $\langle N_i(p)N_j(-p)\rangle=G\big(\delta_{ij}-k_i k_j/k^2\big)/\sigma k^2+...$ corresponding again to nonlocal singularities $\sim\delta^{(1)}(\tau)\log|{\bf x}-{\bf x'}|$, this time in space.

These nonlocal singularities both in time and space are responsible for spurious divergences associated with irregular $1/\omega^2$ and $1/k^{2z}$ terms in the propagators and generically violate a conventional renormalization procedure, so that a subtle step in the proof of renormalizability consists in the search for a special class of gauges in which all propagators are regular.

This class of gauge conditions for generic spacetime dimensionality has been built in \cite{HG} as the following generalization of the relativistic gauge conditions which involve first-order time derivative of the shift functions, spatial derivatives of metric perturbations,
\begin{equation}
    F^i=\dot N^i+\frac{1}{2\sigma}
    \big(\mathcal{O}^{-1}\big)^{ij}\,
    (\partial^kh_{kj}-\lambda\partial_jh),  \label{gauge}
\end{equation}
along with the generically nonlocal in space gauge-fixing operator in the gauge-breaking Lagrangian (\ref{local})
\begin{equation}
    \mathcal{O}_{ij}
    =-(-\Delta)^{2-d}\lp \Delta\delta_{ij}
    +\xi\partial_i\partial_j\rp^{-1}.      \label{O}
\end{equation}
With such a choice the elements of this two parameter family of gauge conditions ($\sigma$ and $\xi$ are free gauge fixing parameters) have the following scaling dimensions
\begin{equation}
    [F^i]=2d-1,\quad[\mathcal{O}_{ij}]=2-2d ,
\end{equation}
which guarantee anisotropic scaling invariance of the full action. The identification of the parameter $\lambda$ with that of the kinetic term of the Ho\v{r}ava action (\ref{genact}) is not accidental --- it allows, in particular, to avoid the cross $Nh$-term in the quadratic part of the gauge-fixed action and thus simplifies the block structure of the full propagator of the theory.\footnote{\label{local_vertices}Moreover, this identification leaves a spatially nonlocal term $\sim\sigma\dot N^i\mathcal{O}_{ij}\dot N^j/2G$ only in the quadratic part of the full gauge-fixed action and does not spoil locality of vertices, which is necessary for correctness of BPHZ renormalization (see below).}

Let us show the regularity of this propagator in $d=3$ case which we will consider in the high-energy limit, so that the coefficients of relevant deformation terms can be set to zero, $\eta=\m_1=\m_2=0$. For that one combines $\LL_{gf}$ with the quadratic Lagrangian (\ref{S2d3}),
$\eta=\m_1=\m_2=0$, and flips the sign of $\nu_{4,5}$ in
consequence of the Wick rotation. Then, a straightforward calculation
yields the non-zero components of propagators,
\beqa
\label{NNniced3}
\langle N^i(p)&&\!\!\!N^j(-p)\rangle=\frac{G k^2}{\s}
(k^2\delta_{ij}-k_ik_j)\,\PP_1(p)
+\frac{\vk^2(1+\xi)k^2}{\s}k_ik_j\,\PP_2(p)\;,\\
\langle h_{ij}(p)&&\!\!\!h_{kl}(-p)\rangle=2G
(\delta_{ik}\delta_{jl}+\delta_{il}\delta_{jk})\PP_{tt}(p)\notag\\
&&-2G\delta_{ij}\delta_{kl}\bigg[\PP_{tt}(p)-\frac{1-\l}{1-3\l}
\PP_s(p)\bigg]\notag\\
&&-2\vk^2(\delta_{ik}\hat k_j\hat k_l+\delta_{il}\hat k_j\hat k_k
+\delta_{jk}\hat k_i\hat k_l+\delta_{jl}\hat k_i\hat k_k)
\big[\PP_{tt}(p)-\PP_1(p)\big]\notag\\
&&+2\vk^2(\delta_{ij}\hat k_k\hat k_l+\hat k_i\hat k_j\delta_{kl})
\big[\PP_{tt}(p)-\PP_s(p)\big]\notag\\
&&+2\vk^2\hat k_i\hat k_j\hat k_k\hat k_l
\bigg[\PP_{tt}(p)+\frac{1-3\l}{1-\l}\PP_s(p)-4\PP_1(p)
+\frac{2\PP_2(p)}{1-\l}\bigg]\;,
\label{hhniced3}
\eeqa
where $\hat k_i=k_i/\sqrt{k^2}$ is a spatial momentum normalized to unit vector and the pole structures are,
\beqa
\label{PTT}
&&\PP_{tt}=\frac{1}{\omega^2+\nu_4k^6}\;,\quad
\PP_s=\Big[\omega^2+\frac{(8\nu_4+3\nu_5)(1-\l)}{1-3\l}\,k^6\Big]^{-1}\;,\\
\label{P1d3}
&&\PP_1=\Big[\omega^2+\frac{k^6}{2\s}\Big]^{-1}\;,\quad
\PP_2=\Big[\omega^2+\frac{(1-\l)(1+\xi)}{\s}\,k^6\Big]^{-1}\;.
\eeqa
The first two structures correspond to the physical TT and scalar modes,
cf. Eqs.~(\ref{dispreld3tt})-(\ref{dispreld3s}), whereas the other two are
gauge-dependent because they involve gauge-fixing parameters $\xi$ and $\sigma$.

The propagator (\ref{NNniced3}) is obviously regular.
For the terms in the last three lines of (\ref{hhniced3}) the situation is subtler. One may worry that the unit vectors entering them contain factors $k$ in the denominator and apparently violate the regularity condition $A_m\neq 0$ in (\ref{reg_int}). However, we observe that the
combinations in the square brackets in these terms vanish at $k=0$,
$\omega\neq 0$. Besides, they depend on the spatial momentum through
$k^6$. This implies that when the worrisome terms are written as
ratios of polynomials, their numerators are at least proportional to
$k^6$, which cancels all powers of $k$ in the denominator. This
cancellation is in fact guaranteed by the regularity of the propagator
$\langle h_{ij}h_{kl}\rangle$ at $k\to 0$, $\omega$--fixed; this, in
turn, follows from the regular structure of the kinetic term of $\LL+\LL_{\rm gf}$ for $h_{ij}$ in this limit.

The ghost sector of the theory can be built by standard rules of the Faddeev-Popov or BRST gauge-fixing procedure. The ghost action is the bilinear combination of the anticommuting ghost fields $c^i$ and $\bar c_j$,
\be
\label{ghostact}
S_{gh}=-\frac1G\int \di\tau\, \di^2x\;\bar c_i\big(\delta^{c}\!F^i\big)\;,
\ee
where $\delta^C\!F^i$ is the linear transformation of gauge-fixing functions --- their linearized foliation preserving diffeomorphism $\delta^f\!F^i$ with the vector parameter $f^i$ identified with the ghost $c^i$. With the finite diffeomorphism given by Eqs.(\ref{x_to_tilde_x}), (\ref{Ftrans}) and (\ref{Ftrans1}) its linearized version, $\tilde x^i=x^i+f^i(x,\tau)$, reads for $h_{ij}$ as the Lie derivative with respect to $f^i$, $\delta^f h_{ij} ={\cal L}_f\gamma_{ij}$, whereas for $N^i$ its ${\cal L}_f N^i$ is also amended by $\dot f^i$,
\beqa
&&\delta^f\! h_{ij}=
\d_i f^k (\delta_{jk}+h_{jk})+\d_j f^k (\delta_{ik}
+h_{ik})+f^k\d_k h_{ij}\;,                     \label{BRSTh}\\
&&\delta^f\! N^i=\dot f^i+f^j\d_j N^i-N^j\d_j f^i\;. \label{BRSTN}
\eeqa

Thus for $d=3$ in the gauge (\ref{gauge})-(\ref{O}) the ghost action reads
\be
\label{ghostactd3}
\begin{split}
S_{gh}=\frac{1}{G}&\int \di \tau \,\di^2x\,\bigg[\,\dot{\bar c}_i\dot c^i
-\frac{1}{2\s}\bar c_i\D^3 c^i
+\frac{1-2\l+2\xi(1-\l)}{2\s}\d_i\bar c_i\D^2\d_j c^j\\
&+\D\dot{\bar c}_i\d_j c^i N^j-\D\dot{\bar c}_i c^j \d_j N^i
+\frac{1}{\s}\D^2\d_j\bar c_i
\Big(\d_{(i}c^k h_{j)k}+\frac12 c^k\d_kh_{ij}\Big)\\
&-\frac{\l(1+\xi)}{\s}\D^2\d_i\bar c_i\Big(\d_k c^l h_{lk}+\frac12 c^l\d_l h\Big)\\
&+\frac{\xi}{\s}\D\d_i\d_j\d_k\bar c_i
\Big(\d_j c^lh_{kl}+\frac12 c^l\d_lh_{jk}\Big)\,\bigg]\;.
\end{split}
\ee
This action is invariant under the anisotropic scaling (\ref{scaling})
with the assignment of zero scaling dimension to the ghosts,
\be
\label{dimghost}
[c^i]=[\bar c_i]=0\;,
\ee
and it gives rise to the ghost propagator also satisfying the regularity condition,
\be
\label{propcc}
\langle c^i(p)\bar c_j(-p)\rangle=\vk^2\delta_{ij}\PP_1(p)
+\vk^2\hat k_i\hat k_j\big[\PP_2(p)-\PP_1(p)\big]\;.
\ee
Similar properties of Ho\v{r}ava gravity models in this class of gauges hold in all higher dimensions, and we are now ready to prove their UV renormalizability.

\subsection{Proof of renormalizability}
For the full set of all quantum fields $\phi=h_{ij},N^i,c^i,\bar{c}_i$ every Feynman graph is characterised by the following set of parameters:\\
$P_{hh}$ --- number of $\langle h_{ij}h_{kl}\rangle$ propagators,\\
$P_{NN}$ --- number of $\langle N^iN^j\rangle$ propagators,\\
$P_{c\bar c}$ --- number of the ghost propagators,\\
$V_{[h]}$ --- number of vertices involving only the $h_{ij}$-fields,\\
$V_{[h]N}$ --- number of vertices with an
  arbitrary number of $h$-legs and a single $N$-leg,\\
$V_{[h]NN}$ --- number of vertices with an
  arbitrary number of $h$-legs and two $N$-legs,\\
$V_{hcc}$ --- number of vertices describing interaction of
  $h_{ij}$ with the ghosts,\\
$V_{Ncc}$ --- number of vertices describing interaction of
  $N^{i}$ with the ghosts,\\
$L$ --- number of loops, i.e. number of independent loop integrals,\\
$l_N$ --- number of external $N$-legs,\\
$T$ --- number of time-derivatives acting on external legs,\\
$X$ --- number of spatial derivatives acting on external legs.

These quantities obey two relations
\begin{align}
\label{Lloops}
&L=P_{hh}+P_{NN}+P_{cc}-V_{[h]}-V_{[h]N}-V_{[h]NN}-V_{hcc}-V_{Ncc}+1\;,\\
\label{Nlines}
&l_N=V_{[h]N}+V_{Ncc}+2V_{[h]NN}-2P_{NN}\;.
\end{align}
The first relation follows from the standard reasoning that out of
$\sum P=P_{hh}+P_{NN}+P_{cc}$ original integrals over frequencies and momenta
$\sum V-1=V_{[h]}+V_{[h]N}+V_{[h]NN}+V_{hcc}+V_{Ncc}-1$ of them
are removed by the $\delta$-functions
at the vertices (one $\delta$-function remains as an overall factor
multiplying the whole diagram). The second relation is obtained by
counting the $N$-legs. Indeed,
each vertex of the type $V_{[h]N}$ or $V_{Ncc}$
brings one $N$-leg, whereas the vertex $V_{[h]NN}$ brings two;
every $\langle N^iN^j\rangle$-propagator absorbs two
legs; the remaining $N$-legs are external.

The numbers of time and space derivatives in vertices are given in the following table
\begin{table}[h!]
\centering
\begin{tabular}{|c|c|}
\hline
\; vertex\; &\; \# of vertex derivatives  $\vphantom{L^{L^{L^I}}_{L_{L_I}}}$\; \\ \hline
$V_{[h]}$      & $\partial_x^{2d} \vphantom{L^{L^{L^I}}_{L_{L_I}}}$      \\ \hline
$V_{[h]N}$ & $\partial_\tau\partial_x \vphantom{L^{L^{L^I}}_{L_{L_I}}}$ \\ \hline
$V_{[h]NN}$    & $\partial_x^{2} \vphantom{L^{L^{L^I}}_{L_{L_I}}}$        \\ \hline
$V_{hc\bar c}$ & $\partial_x^{2d} \vphantom{L^{L^{L^I}}_{L_{L_I}}}$       \\ \hline
$V_{Nc\bar c}$ & $\partial_\tau\partial_x \vphantom{L^{L^{L^I}}_{L_{L_I}}}$  \\ \hline
\end{tabular}
\end{table}

The scaling dimensionality of any block of the full propagator in momentum space is obviously related to dimensionalities of relevant fields in the coordinate space $[\braket{\phi_1(p)\phi_2(-p)}]_p=[\phi_1]_x+[\phi_2]_x-2d$, so that from the dimensions of fields in $x$-space,
\begin{equation}
    [h]_x=0,\quad [N]_x=d-1,\quad [c]_x=[\bar c]_x=0,
\end{equation}
one finds the scaling dimensions of various blocks of their two-point momentum-space propagators,
\begin{equation}
    [\braket{hh}]_p=-2d,\quad[\braket{NN}]_p=-2,\quad[\braket{c\bar c}]_p=-2d.
\end{equation}
Hence, the degree of divergence of any Feynman diagram equals
\begin{equation}
\begin{split}
    \mathcal{D}_{\mathrm{div}}&=2dL-2dP_{hh}-2P_{NN}-2dP_{c\bar c}\\
    &+2dV_{[h]}+(d+1)V_{[h]N}+2V_{[h]NN}+2dV_{hc\bar c}+(d+1)V_{Nc\bar c}\\
    &-dT-X=2d-dT-X-(d-1)\ell_N,
\end{split}
\end{equation}
where the first line is contributed by loop momenta integration measure and the the full set of propagators, the second line is a contribution of vertex scalings and the third line corresponds to the reduction of the total scaling of degree of divergence due to time and space derivatives on the external lines.

Using the above relations (\ref{Lloops}) and (\ref{Nlines}) we get a remarkable expression for $\mathcal{D}_{\mathrm{div}}$ which is independent of the internal structure of the diagram,
\begin{equation}
    \mathcal{D}_{\mathrm{div}}=2d-dT-X-(d-1)\ell_N.
\end{equation}
We see that $D_{div}$ is negative for diagrams with more than 2 time- or $2d$ space-derivatives on external legs. Therefore, only diagrams with at most 2 time- and $2d$ space-derivatives on the external lines must be renormalized. The corresponding counterterms are polynomial in external frequencies and momenta and hence local in position space. Again, they have no more than 2 time- or $2d$ space-derivatives acting on the metric $h_{ij}$. In other words, their scaling dimension is less or equal four. If we further assume that the divergent parts of the diagrams respect the local foliation-preserving diffeomorphisms --- this will be discussed in the next section, it follows that the counterterms must have the same form as the terms already present in the action (\ref{genact}), (\ref{Vd2}). This amounts to renormalizability \cite{HG}.

\subsection{Non-projectable models \label{Non-projectable}}
Non-projectable Ho\v rava gravity models have extra symmetry $t\mapsto \tilde t( t)$ and generic space-inhomogeneous lapse function $N=N(t,\textbf{x})\neq1$, which leads to extra contributions in the potential term of the action (\ref{genact}). For illustration of the general situation we take the model in $(2+1)$-dimensions \cite{Sotiriou:2011dr} which is technically much simpler than its $(3+1)$-dimensional counterpart. In this case the potential contains 10 inequivalent terms,
\be
\begin{split}
\label{Vd2NP}
{\cal V}=&2\Lambda-\eta R-\a a_ia^i+\m R^2+\rho_1\D R+\rho_2 R a_ia^i
\\
&+\rho_3 (a_ia^i)^2+\rho_4a_ia^i\nabla_ja^j+\rho_5(\nabla_j a^j)^2
+\rho_6\nabla_ia_j\nabla^ia^j\;,
\end{split}
\ee
where
\be
\label{ai}
a_i=\d_i\ln N
\ee
is the ``acceleration'' variable which is invariant under the reparameterizations of time, see Eqs.~(\ref{Ftrans}). Again tuning the cosmological constant $\Lambda$ to zero  and expanding around flat spacetime, one obtains the quadratic action,
\be
\label{S2NP}
\begin{split}
S^{(2)}_{d=2,\,\rm n-p}=\frac1G\int \di t\,\di ^2x\,&\bigg[
\!-\frac{1}{2}(\dot v_i-u_i)\D(\dot v_i-u_i)
+\frac{\dot\psi^2}{4}+\frac{1}{4}(\dot E-2\D B)^2\\
&-\frac{\l}{4}(\dot\psi+\dot E-2\D B)^2-\m(\D\psi)^2\\
&-\eta\phi\D\psi-\a\phi\D\phi
+\rho_1\phi\D^2\psi-(\rho_5+\rho_6)\phi\D^2\phi\bigg],
\end{split}
\ee
where in addition to the projectable case (\ref{S2d2}) one gets the contribution of the fluctuation of the lapse $\phi\equiv N-1$ --- a new variable devoid of the kinetic term. Exclusion of this variable by its variational equation -- the second class constraint -- leads to the
propagation of a single scalar degree of freedom with the dispersion
relation non-polynomial in the momentum,
\be
\label{disprelNP}
\omega^2=\left(\frac{1-\l}{1-2\l}\right)
\frac{\eta^2k^2+(4\a\m+2\eta\rho_1)k^4
+\big(\rho_1^2-4\m(\rho_5+\rho_6)\big)k^6}{\a-(\rho_5+\rho_6)k^2}\;.
\ee
In contrast to the projectable case, this dispersion relation is
linear at low $k$, $\omega^2=k^2\eta^2(1-\l)/\a(1-2\l)$, but at large momenta it respects the anisotropic scaling (\ref{scaling}),
\be
\label{disprelNPhighk}
\omega^2=\frac{1-\l}{1-2\l}\bigg(4\m-\frac{\rho_1^2}{\rho_5+\rho_6}\bigg)
\; k^4\;.
\ee
The mode has positive energy and is stable at all momenta for
an appropriate choice of parameters $G>0$, $\l<1/2$ or $\l>1$, $\a>0$, $(\rho_5+\rho_6)<0$ and $4\m>\rho_1^2/(\rho_5+\rho_6)$.

If one considers the UV behavior of the model, which amounts to setting $\eta=\a=0$, and considers the gauge-fixing procedure with (\ref{gauge}) and (\ref{O}), then the resulting propagators will have irregular terms in their $\langle\phi\phi\rangle$ and $\langle\phi h_{ij}\rangle$ sectors \cite{HG}. They cannot be removed by any gauge choice and correspond to the
instantaneous interaction present in the theory
\cite{Blas:2010hb,Blas:2011ni}. We conclude that the correlators of
the lapse contain genuinely non-local terms violating sufficient conditions of the renormalizability of the theory, derived above. This casts serious doubt on renormalizability of non-projectable Ho\v{r}ava model. However, regularity of propagators is only a sufficient rather than necessary condition of renormalizability, and there might be subtle mechanisms of cancellation of irregular UV divergences caused by these terms.

The source of these terms is, actually, the presence of second-class constraints in the canonical formalism of non-projectable model \cite{Belorin2020,Devecioglu,Belorin2021,Belorin2022a}. It should be emphasized that with this type of constraints direct use of Faddeev-Popov gauge fixing \cite{Faddeev-Popov} in Lagrangian formalism is insufficient to produce correct Feynman diagrammatic technique -- it should be based on the canonical BFV quantization \cite{BFV1,BFV2,BFV3} of systems subject to a combination of first-class and second-class constraints. It incorporates nontrivial, time-local $\sim\delta^{(1)}(0)$, path integral measure \cite{Fradkin,Senjanovic,Fradkin-Fradkina} and Dirac brackets in canonical phase space of the theory. Eradication of the second-class constraints by directly solving them as well as calculation of Dirac brackets generically result in spatial nonlocality of the action which again compromises local renormalization technique. Interestingly, due to peculiarities of the Hor\v{r}ava model this type of nonlocality was circumvented in \cite{Belorin2022} and, moreover, for the remaining irregular propagator terms the mechanism of their cancellation was observed. In particular, for the (2+1)-dimensional model the power divergent set of contributions of irregular terms $\sim\!\text{\small $\int$} d\omega$ were shown to be cancelled by the contribution of the local measure $\sim\!\delta^{\scriptstyle (1)}(0)=\text{\small $\int$} d\omega/2\pi$. This opens serious prospects for non-projectable Ho\v{r}ava model that could have served as the only known at present candidate for local, unitary, renormalizable gravity theory compatible with general relativity in IR domain \cite{Blas:2010hb}.

\section{BRST structure of renormalization and covariance of counterterms \label{BRST}}

We also must provide the gauge invariance of the counterterms. In
the perturbative expansion around flat spacetime, considered so far, gauge invariance is actually {\em not} preserved. The way to
proceed would be to exploit the BRST symmetry of the gauge-fixed
action to constrain the structure of counterterms, similar to the
analysis of \cite{Stelle:1976gc}. Even more efficient is to adopt the background field formalism and the method of background covariant gauges \cite{DeWitt_covariant,Veltman:1975vx,Abbott:1981ke} where the gauge invariance becomes manifest.

BRST structure of the renormalization in such gauges, which is supposed to lead to covariant counterterms and reduce to the renormalization of the coupling constants in the original action (\ref{genact}), requires extension of known results for Lorentz invariant theories to Ho\v{r}ava gravity. Such an extension is possible \cite{BRST} within a class of theories with a generic {\em closed} algebra of {\em irreducible} gauge generators which are {\em linear} in the quantum fields $\varphi$. This extension runs via the inclusion into the conventional BRST operator $Q$ the background field $\phi$ and its BRST partner along with a special choice of the gauge fermion $\varPsi_{\rm ext}[\varPhi,\varPhi^*]$ which depends on the full set of quantum fields $\varPhi=(\varphi,c,\bar c,b)$ (including together with $\varphi$ the BRST ghosts $c$, $\bar c$ and Lagrange multipliers $b$) and their antifields $\varPhi^*$, the latter playing the role of the sources of the BRST transformations of the full set of $\varPhi$,
 \begin{eqnarray}
Q\Rightarrow Q_{\rm ext},\quad
\varPsi\Rightarrow\varPsi_{\rm ext}[\varPhi,\varPhi^*].
\end{eqnarray}
The resulting generating functional $W[J,\varPhi^*]$ --- the functional of the sources $J$ of the quantum fields $\varPhi$ and their antifields $\varPhi^*$, which is given by the path integral
 \begin{eqnarray}
\exp\Big(-\frac1\hbar W[J,\varPhi^*]\Big)=\int D\varPhi\,
e^{-\big(S[\varphi]+
Q_{\rm ext}\varPsi_{\rm ext}[\varPhi,\varPhi^*]+J\varPhi\big)/\hbar},
\end{eqnarray}
satisfies well known Slavnov-Taylor identities and also solves special Ward identities. The latter hold provided the gauge fermion is built of the so-called {\em background covariant gauge conditions} which make the gauge fermion invariant under the simultaneous gauge transformations of both the quantum field $\varphi$ and its background counterpart $\phi$.

Application of Slavnov-Taylor and Ward identities to the divergent part of the effective action runs perturbatively in $\hbar$ via the study of the cohomologies of the nilpotent BRST operator $Q_{\rm ext}$. It yields the needed BRST structure --- the overall effect reduces to a simultaneous local renormalization of the action $S[\varphi]$ by gauge invariant counterterms of the original gauge fields $\varphi$ and the renormalization of the gauge fermion $\varPsi_{\rm ext}[\varPhi,\varPhi^*]$ which also gets quantum corrections,
\begin{eqnarray}
&&S[\,\varphi\,]\to S[\,\varphi\,]+\Delta_\infty S[\,\varphi\,],\\ &&\varPsi_{\rm ext}[\,\varPhi,\varPhi^*]\to
\mbox{\boldmath$\varPsi$}_{\rm ext}[\varPhi,\varPhi^*]
+\Delta_\infty\mbox{\boldmath$\varPsi$}_{\rm ext}[\varPhi,\varPhi^*].
\end{eqnarray}

This BRST structure of renormalization is achieved by means additional renormalization of quantum fields, which turns out to be a (generically nonlinear) anti-canonical transformation generated by the gauge fermion $\varPsi_{\rm ext}$ itself \cite{BRST}. It is important that the uncontrollably complicated renormalization of the gauge fermion, indiscriminately depending on all quantum fields $\varPhi$, is immaterial from the viewpoint of physical applications, because anyway the generating functional of physical amplitudes is gauge independent onshell,
$\delta_\varPsi W\,|_{\,J=\varPhi^*=0}=0$. Remarkable feature of this scheme is that it applies not only to perturbatively renormalizable theories, but also to effective field theories below their cutoff \cite{BRST}. All this justifies the physically invariant scope of these results and their applications, in particular, to Ho\v{r}ava gravity models.

In the context of Ho\v{r}ava gravity the construction of background covariant gauges starts with the decomposition of the full set of gauge fields $\varphi=(\gamma_{ij},N^i)$ into their background $\phi=(g_{ij},{\cal N}^i)$ and quantum fluctuations $(h_{ij}$, $n^i)$:
\be
\label{decomp}
\gamma_{ij}=g_{ij}+h_{ij}~,~~~~N^i={\cal N}^i+n^i\;.
\ee
The background covariant gauge conditions for these fluctuations are just the  covariantization of all formulas of the previous sections. Instead of
(\ref{gauge}) and (\ref{O}) we write,
\beqa
\label{Fcovar}
&&F^i=D_t n^i+\frac{1}{2\s}(\OO^{-1})^{ij}\big(D_k h^k_j
-\frac{\l}{2\s}D_j h\big)\;,\\
\label{Ocov}
&&\OO_{ij}=\Big[g^{ij}(-\Delta)^{d-1} - \xi D^i(-\Delta)^{d-2}D^j\Big]^{-1}\;.
\eeqa
where
\be
\label{timecov}
D_t n^i=\dot n^i-{\cal N}^k D_k n^i+n^k D_k{\cal N}^i
\ee
is the covariant time-derivative and $D_i$ are the covariant derivatives  conserving the background metric $g_{ij}$, $\Delta=g^{ij}D_i D_j$ is the respective covariant Laplacian, all indices are raised and lowered by $g^{ij}$ and $g_{ij}$, and $h=h_{ij}g^{ij}$. Lack of commutativity of $D_i$ explains the operator ordering in the definition (\ref{Ocov}) of the symmetric operator $\OO_{ij}$. The gauge-fixing action is still given by the Lagrangian (\ref{local}) that must be integrated over the spacetime with the covariant measure $\int \di \tau\, \di^d x\sqrt{g}$, $g=\det g_{ij}$. Finally, for the gauge transformations (\ref{BRSTh})-(\ref{BRSTN}) their covariantization reduces to identically rewritring them in the form
\beqa
&&\delta^f\! h_{ij}=
D_i f_j +D_j f_i,\quad f_i=g_{ij}f^j\;,                     \label{BRSThcov}\\
&&\delta^f\! n^i=\dot f^i+f^j D_j n^i-n^j D_j f^i\;. \label{BRSTNcov}
\eeqa

One should be worried at this point that the gauge-fixing Lagrangian
depends on the background fields in a non-local manner which can
compromise the locality of counterterms (see footnote \ref{local_vertices}). To resolve this issue, we observe that the non-local operator $\OO_{ij}$
actually cancels everywhere in the gauge-fixing action, except the
kinetic term for the shift,
\be
\label{nkin}
S_{n,\,kin}[\,n\,]=\frac{\s}{2G}\int \di\tau\,\di^dx\,\sqrt{g}\,
D_t n^i\OO_{ij} D_t n^j\;.
\ee
The latter is cast in the local form by introducing an auxiliary field
$\pi_i$,
\be
\label{nkinpi}
S'_{n,\,kin}[\,\pi,n\,]=\frac{1}{G}\int \di\tau\,\di^dx\,\sqrt{g}\,\bigg[
\frac{1}{2\s}\pi_i(\OO^{-1})^{ij}\pi_j-i\pi_i D_tn^i\bigg]\;.
\ee
Taking the Gaussian path integral over $\pi_i$ reproduces
(\ref{nkin}). Note that we have introduced an imaginary coefficient
in front of the second term in (\ref{nkinpi}) in order to preserve the
positivity of the quadratic term\footnote{Strictly speaking, this
  argument applies when the operator $\OO_{ij}$ is positive-definite, but a possible lack of positivity does not affect the
  perturbative considerations.}.
Note that $\pi_i$ enters in the action as a canonically conjugate momentum
for the shift perturbations $n^i$. From this perspective, the presence
of an imaginary part in (\ref{nkinpi}) is not surprising, because the
imaginary part associated with the canonical form always appears when the Euclidean action is written in terms of canonical variables.

It is instructive to work out how the introduction of $\pi_i$ affects
the measure in the path integral. Let us make a step backward and
recall that the gauge-fixing Lagrangian (\ref{local}) arises as a
result of smearing the delta-function type gauge-fixing condition $F^i=f^i$ with the
weighting functional
\be
\label{weighting}
\big(\Det \OO_{ij}\big)^{1/2}\int Df \,\exp\Big[-\frac\s{2G}\int \di\tau\,\di^d x \sqrt{g}\,f^i\OO_{ij}f^j\,\Big]
\ee
inserted in the partition function of the theory. Notice the square
root of the functional determinant of the operator $\OO_{ij}$
in the prefactor which ensures the
correct normalization. Thus, before
introducing $\pi_i$ the partition function has the form,
\be
\label{partit1}
Z=\big(\Det \OO_{ij}\big)^{1/2}\int Dn\,Dh\,Dc\,D\bar c
\exp\big[-(S_{n,\,kin}+\ldots)\big]\;,
\ee
where ellipsis stands for the local contributions in the action. The
introduction of $\pi_i$ not only makes the action local, but also
absorbs the determinant from the prefactor, which follows
from the relations
\begin{align}
&e^{-S_{n,\,kin}[n^i]}=\big(\Det \OO_{ij}\big)^{-1/2}
\int D\pi\,e^{-S'_{n,\,kin}[\pi_j,n^i]}\;,\notag
\end{align}
so that the final expression for the partition function reads
\be
\label{partit2}
Z=\int D\pi\,Dn\,Dh\,Dc\,D\bar c \exp\big[-(S'_{n,\,kin}+\ldots)\big]\;.
\ee
Curiously, the introduction of $\pi_i$ makes the integration measure
in the path integral flat (Liouville like), which further supports the identification of $\pi_i$ as the canonically conjugate momentum to~$n^i$.

Finally, we have to check that the introduction of $\pi_i$ does
not spoil the regular structure of the propagators. This is easy to see from the fact that additional blocks of the full propagator $\langle \pi_i(p)n^j(-p)\rangle$ and $\langle \pi_i(p)\pi_j(-p)\rangle$ are also regular and compatible with the scaling dimension $[\pi_i]=1$. As a consequence, the reasoning of previous sections remains true with the field $\pi_i$ included into consideration.

\section{Asymptotic freedom in $(2+1)$-dimensional Ho\v rava gravity \label{AF2+1}}

Here we show that $(2+1)$-dimensional Ho\v{r}ava gravity is asymptotically free in UV limit \cite{2+1}. Its action in the UV domain,
\begin{equation}
    S=\frac{1}{2G}\int dt\,d^2x\,\sqrt{\gamma}
    \lp K^2_{ij}-\lambda K^2+\mu R^2\rp,    \label{2+1}
\end{equation}
includes three coupling constants $G$, $\lambda$ and $\mu$. However, only two their combinations are {\em essential} \cite{Weinberg:1980gg} in the sense that their renormalization does not depend on the choice of the gauge. The gauge variation induces the transformation of the effective action by terms vanishing on shell, that is on effective equations of motion of the theory \cite{DeWitt_covariant,Kallosh}. Since in background covariant gauges the UV divergent part is local and gauge invariant, such a change in the one-loop order can only be of the form
\begin{eqnarray}
    \varGamma^{\mathrm{div}}_{\text{1-loop}}
    \rightarrow&\;\varGamma^{\mathrm{div}}_{\text{1-loop}}&+\varepsilon\int \di\tau\,d^2x\,\frac{\delta S}{\delta \gamma_{ij}}\gamma_{ij}\nonumber\\
    =&\;\varGamma^{\mathrm{div}}_{\text{1-loop}}&+
    \frac\varepsilon{2G}\int dt\,d^2x\,\sqrt{\gamma}
    \lp K^2_{ij}-\lambda K^2-\mu R^2\rp,      \label{var_gauge}
\end{eqnarray}
where $\varepsilon$ parameterizes the gauge variation. This is because other combinations of equations of motion are either noncovariant or do not have a needed dimension.\footnote{Simple derivation of the equality here follows from the fact that under the metric rescaling by a global parameter $a$, $\gamma_{ij}\to a\gamma_{ij}$, the kinetic term of the action gets rescaled linearly in $a$ while the potential term gets rescaled by $1/a$.} Such a variation of the UV counterterm implies the following change in the coupling constants of the theory,
\begin{equation}
    \delta_{\varepsilon}G=-2G^2\varepsilon,\quad \delta_{\varepsilon}\lambda=0,\quad \delta_{\varepsilon}\mu=-4G\mu\varepsilon,
\end{equation}
and implies that only two their combinations are gauge independent. One of them is the original coupling $\lambda$ and another one is
\begin{equation}
    {\cal G}\equiv\frac{G}{\sqrt\mu},\quad\delta_{\varepsilon}{\cal G}=0.
\end{equation}

To perform renormalization of the theory we use the background field formalism of the previous section and calculate the divergent part of the one-loop effective action $\varGamma^{\mathrm{div}}_{\text{1-loop}}[g_{ij},{\cal N}^i]$ at zero background shift functions ${\cal N}^i=0$ and at the background metric $g_{ij}=\delta_{ij}+H_{ij}$, which is close to flat space, by perturbations in $H_{ij}$. Due to the invariance of the effective action, its divergent part has the structure of the classical action (\ref{2+1}) whose expansion begins with the terms quadratic in $H_{ij}$ . The UV counterterms and relevant $\beta$-functions are then found by studying how the two-point functions of $H_{ij}$ -- the coefficients of expansion of the effective action in $H_{ij}$ -- are renormalized after integrating out the quantum fluctuations $h_{ij}$ and $n^i$ (note that $h_{ij}$ and $n^i$ are quantum fluctuations on top of the perturbed background that should be discerned from background perturbations). The renormalization of $G$ is then extracted from the terms $\dot{H}_{ij}^2\sim K_{ij}^2$, while the one of $\lambda$ comes from $\dot{H}^2 \sim K^2$. For the renormalization of $\mu$ we can use any of the three structures $\d_i\d_j H^{ij}\Delta H$, $(\d_i\d_j H^{ij})^2$ or $(\Delta H )^2$ contributing to the $R^2$-potential of (\ref{2+1}). This explains why we do not need the effective action at nonzero ${\cal N}^i$ or the diagrams with external ${\cal N}^i$-lines.

\begin{figure}
\begin{center}
\includegraphics[scale=1]{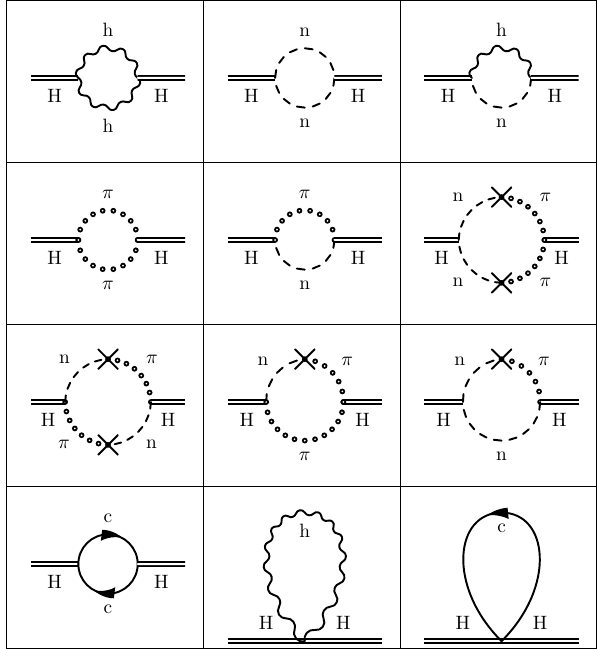}
\end{center}
\caption{Feynman diagrams (bubbles and fishes) for the two point function of $H_{ij}$. The cross represents the mixed propagator $\langle n^i\pi^j\rangle $.}\label{fig:diagrams}
\end{figure}

The propagators of quantum fields $h_{ij}$ and $n_i$ considered above, the momentum $\pi_i$ (introduced in (\ref{nkinpi})), and the ghosts $\bar{C}^i$ and $C^i$ are particularly simple in the gauge
\begin{align}\label{eq:sigmaxi}
\sigma=\frac{1 - 2 \lambda}{8 \mu (1 - \lambda)}, \qquad \xi=-\frac{1 - 2 \lambda}{2 (1 - \lambda)},
\end{align}
where they are all proportional to the propagator of the physical
scalar mode (see Eq.(\ref{dispreld2}) for the dispersion relation of $d=2$ model in Lorentzian spacetime),
\begin{equation}
 \begin{split}
&{\cal P}_{\rm s}(\omega,p)=\left[\omega^2 +4\mu \frac{1-\lambda}{1-2\lambda}\;p^4\right]^{-1}. \label{eq:poles}
\end{split}
\end{equation}
The vertices required for the one-loop calculation can be found by
expanding the total gauge-fixed action up to second order in the background field $H_{ij}$. The diagrams which give rise to logarithmic divergences are shown in Fig.~\ref{fig:diagrams}.


The computation is simplified by considering the renormalization of
$\{{ G},\lambda\}$ and of $\mu$ separately. This can be done by evaluating the quadratic part of the effective action $\Gamma[\delta_{ij}+H_{ij}]$ on
time- or space-dependent backgrounds which correspond respectively to diagrams with vanishing spatial momenta or frequency in external legs.
Thus, $\{{ G},\lambda\}$-renormalization follows from the logarithmically divergent diagrams carrying only external frequency $\Omega$ at vanishing external momentum $P_i$, whereas $\mu$-renormalization originates from those with only external momentum at vanishing $\Omega$.

The typical loop integral over internal momentum and frequency
has the form,
\begin{align}
\label{eq:inte}
\int \frac{\di\omega\,  \di^2q}{(2\pi)^3} \; \omega^{2a} q^{2b}\prod_I {\cal P}_{s} (\omega+\Omega_I,\,q+P_I),
\end{align}
with constant $a$, $b$ and $\{\Omega_I,P_I\}$ -- the relevant external frequencies and two-momenta. Logarithmically  divergent contributions proportional to $\Omega^2$ or $P^4$, which renormalize the
terms of the bare action (\ref{2+1}) follow from the Taylor expansion of
the integrand of \eqref{eq:inte} up to the desired order in external
frequency or momentum, such that the final integrands all acquire the general form
\begin{align}
\label{eq:integrand}
{\cal I}[a,b,A]=\omega^{2a} q^{2b} \big({\cal P}_s(\omega,q)\big)^A
=\frac{\omega^{2a} q^{2b}}{\Gamma(A)}\int_0^\infty \di s \; s^{A-1}\; e^{-s ({\cal P}_s(\omega,q))^{-1} },
\end{align}
with $A$ being a constant power. The integral over frequency and momentum in \eqref{eq:inte} can then be expressed in terms of the $\Gamma$-functions. In this ``proper time'' representation logarithmic UV divergences appear as the integral divergent at $s=0$, $\int_0^\infty \di s/s$, which when regulated by the UV cutoff $\Lambda_{\rm UV}$ reads as
\begin{align}
\int_0^\infty\frac{\di s}s\mapsto
\log\left(\frac{\Lambda_{\rm UV}^4}{k_*^4}\right)\;, \label{cutoff2}
 \end{align}
where $k_*$ is a subtraction point. Here we have taken into account that the proper time parameter $s$ has scaling dimension $4$.

As a result the UV finite {\em renormalized} coupling constants $G$ and $\nu_a=\{\lambda,\mu\}$, $a=1,2$, express in the one-loop approximation in terms of the bare (divergent) couplings $G_0,\nu_a^0$ via the following equations
\begin{equation}
\frac1{2G} =  \frac1{2G_0}
+ C_G\ln\frac{\varLambda_{\rm UV}^2}{k_*^2},\quad
\frac{\nu_a}{2G}=
\frac{\nu^0_a}{2G_0}
+ C_{\nu_a}\ln\frac{\varLambda_{\rm UV}^2}{k_*^2},     \label{Gnu_0}
\end{equation}
where $C_G$ and $C_{\nu_a}$ are some independent of $G$ (and $G_0$) coefficient functions of $\lambda$ and $\mu$ -- a primary goal of one-loop calculations. Then, the full set of beta-functions of all renormalized couplings, defined according to standard rules of renormalization group theory as the derivatives with respect to the running scale $k_*$, reads
\begin{equation}
\beta_G\equiv \frac{dG}{d\ln k_*}
= 4 G^2 C_G,\quad
\beta_{\nu_a}\equiv \frac{d\nu_{a,\,{\rm ren}}}{d\ln k_*}
= -4 G C_{\nu_a}
+\nu_a\frac{\beta_G}G.         \label{beta_Gnu}
\end{equation}
Calculation of $C_G$ and $C_{\nu_a}$ then gives \cite{2+1}
\begin{eqnarray}
&&\beta_{\lambda}
=\frac{15-14\lambda}{64\pi}
\sqrt{\frac{1-2\lambda}{1-\lambda}}\; {\cal G}, \\
&&\beta_{\mu}
=\frac{30-73\lambda+42\lambda^2}{32\pi(1-\lambda)^{3/2}\sqrt{1-2\lambda}}
G\sqrt\mu,\;\;
\beta_G
=-\frac{30\lambda-23}{32\pi
\sqrt{(1-2\lambda)(1-\lambda)}}\frac{G^2}{\sqrt\mu}. \label{mu_G}
\end{eqnarray}
The last two beta-functions separately do not make much sense, because their couplings are not essential and depend on the choice of gauge, but their combination yields the beta-function of essential ${\cal G}=G/\sqrt\mu$,
\begin{eqnarray}
&&\beta_{\cal G}
=-\frac{(16-33\lambda +18\lambda^2)}{64\pi (1-\lambda)^2}
\sqrt{\frac{1-\lambda}{1-2\lambda}}\; {\cal G}^2,     \label{cal_G}
\end{eqnarray}
which is indeed gauge independent as it can be checked by the calculation in the alternative gauge. Namely, it was checked in the background gauge with $\xi=0$ (and with $\sigma$ as in (\ref{eq:sigmaxi})) and also outside of the family (\ref{Fcovar})-(\ref{Ocov}), in the conformal gauge,
$h_{ij}=e^{2\phi}\gamma_{ij}$, which is possible in two spatial
dimensions.\footnote{In the degenerate (delta-function type) conformal gauge the beta-functions of $G$ and $\mu$ are different from (\ref{mu_G}),
 \begin{equation*}
\beta_{\mu}
=\frac{2-7\lambda+6\lambda^2}{32\pi(1-\lambda)^{3/2}\sqrt{1-2\lambda}}
\,G\sqrt\mu,\quad
\beta_G
=-\frac{6\lambda-7}{32\pi
\sqrt{(1-2\lambda)(1-\lambda)}}\;\frac{G^2}{\sqrt\mu},
\end{equation*}
but they lead to the same $\beta_{\cal G}$ given by (\ref{cal_G}).
} Furthermore, it matches with the results of
\cite{Griffin:2017wvh}.

\begin{figure}[h]
\vspace{0.2cm}
\centerline{\includegraphics[width=0.46\textwidth]{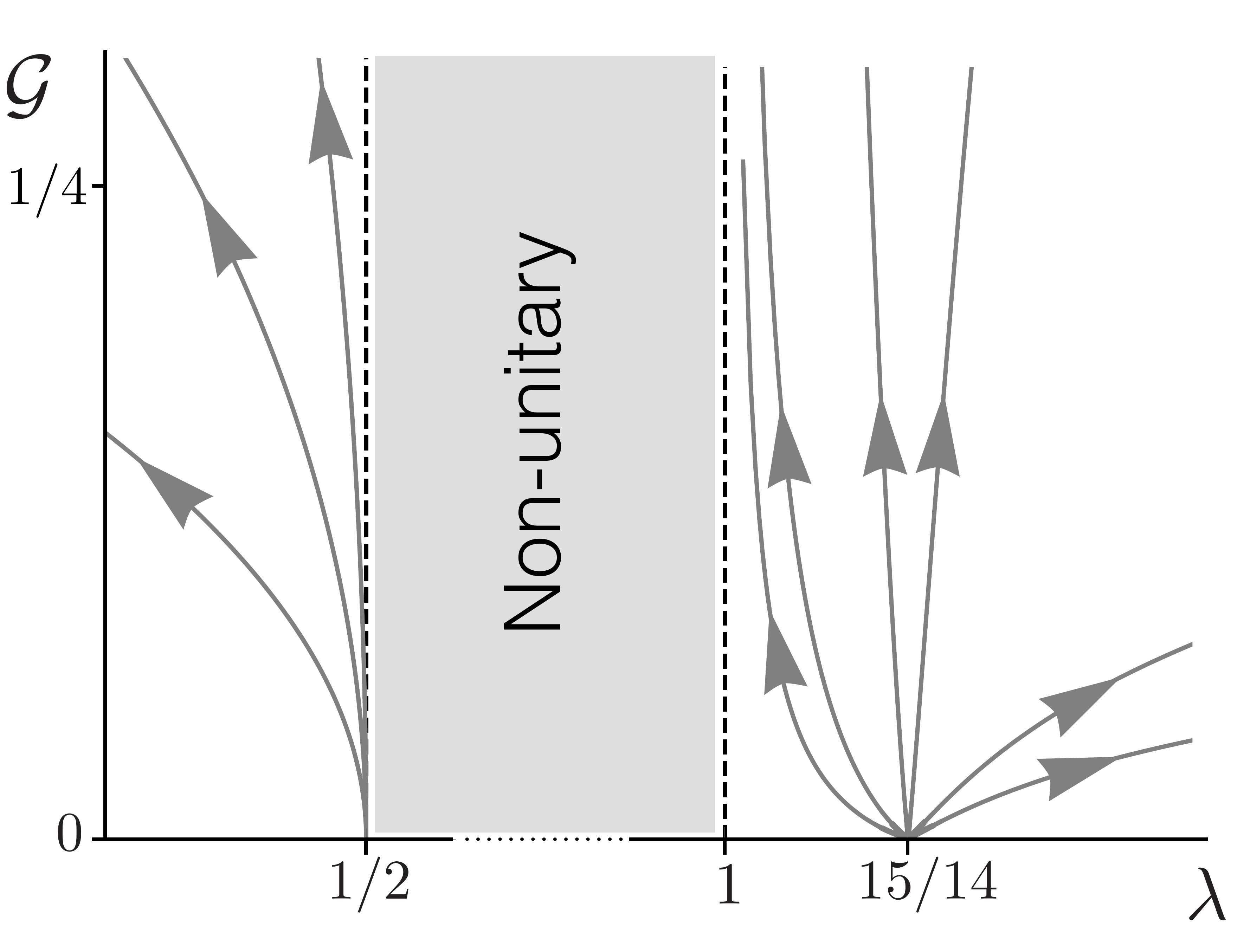}}
\caption{RG flow of the couplings in
  $(2+1)$-dimensional Ho\v rava gravity. The arrows show the direction of
  the flow towards the infrared.}
\label{flow}
\vspace{-0.2cm}
\end{figure}

The structure of the RG flow generated by beta-functions in the regions of unitarity, $\lambda<1/2$ and $\lambda>1$, is shown in
Fig.~\ref{flow}. The theory possesses two UV fixed points:
$(\l,{\cal G})=\left(1/2,0\right)$ and $(\l,{\cal G})=\left(15/14,0\right)$.
The first fixed point is located at the boundary of the allowed region and corresponds to strong coupling, as is clear from the singularity in \eqref{cal_G}. However, the structure of $\beta_{\cal G}$ suggests that the actual expansion parameter in the limit $\lambda\to 1/2$ is $\tilde{\cal
G}={\cal G}/\sqrt{1-2\l}$, with the $\beta$-function $\beta_{\tilde {\cal G}}=-(1-2\l)^2\tilde{\cal G}^2/64\pi (1-\l)^{3/2}$. It vanishes at $\lambda\to 1/2$, so that $\tilde {\cal G}$ freezes at a constant value in the UV --- at the one-loop level there is a family of UV fixed points parameterized by the asymptotic value of $\tilde {\cal G}$. The status of this fixed-point family can be clarified only by taking into account contributions from higher order and matter loops.

The second UV fixed point $(\l,{\cal G})=\left(15/14,0\right)$ is regular and asymptotically free. In the infrared (IR), the RG trajectories either go to $\l\to +\infty,~{\cal G}\to +\infty$, or to $\l\to 1^+,~{\cal G}\to+\infty$. The latter behavior naively corresponds to the relativistic limit of the theory. However, to conclude whether the theory really flows or not to GR requires a non-perturbative analysis as in the IR the system enters into the
strong-coupling regime, which is typical for asymptotically free theories. Thus, the RG flow possesses an asymptotically-free fixed point in the UV,
which establishes this model as a $2+1$ dimensional perturbatively
UV-complete theory with a non-trivial propagating gravitational degree of freedom.

\section{One-loop beta functions of (3+1)-dimensional Ho\v rava gravity\label{3+1}}
Physically the most interesting is, of course, the case of $(3+1)$-dimensional theory. Its UV behavior is determined by the action
\begin{eqnarray}
S= \frac{1}{2G} \int d\tau\, d^3x \,\sqrt{\gamma}&\big(&K_{ij}K^{ij}-\lambda K^2 + \nu_1 R^3+\nu_2 R R_{ij}R^{ij}
+\nu_3 R^i_j R^j_k R^k_i\nonumber\\
&& + \nu_4 \nabla_i R \nabla^i R
+ \nu_5 \nabla_i R_{jk} \nabla^i R^{jk}\big),           \label{action31}
\end{eqnarray}
where we retain only its marginal operators. As in $(2+1)$-dimensional theory, here only a subset of combinations of seven coupling constants, $(G,\lambda,\nu_a)$, $a=1,...5$, forms essential couplings.

Similarly to (\ref{var_gauge}) the $\varepsilon$-variation of the gauge induces the shift of the one-loop divergent part of the effective action by the integrated trace of $\gamma_{ij}$-equations of motion,  corresponding to the rescaling of the three-metric by a global parameter, $\gamma_{ij}\to a\gamma_{ij}$. Under this rescaling kinetic and potential terms of the action scale respectively as $a^{\pm3/2}$, so that the variation of the one-loop counterterm is again proportional to the difference of the kinetic and potential terms of the classical Lagrangian \cite{towards}. This means that the renormalized couplings vary under the change of the gauge as $\delta_\varepsilon G=-2G^2\varepsilon$, $\delta_\varepsilon\l=0$ and $\delta_\varepsilon\nu_a=-4G\nu_a\epsilon$, whence the set of gauge independent couplings can be chosen as ${\cal G}=G/\sqrt{\nu_5}$, $\lambda$ and $\nu_a/\nu_5$. More useful choice is
\be
{\cal G} = \frac{G}{\sqrt{\nu_5}},\quad \lambda,
\quad u_s=\sqrt{\frac{1-\l}{1-3\l}\left(\frac{8\nu_4}{\nu_5}+3\right)},
\quad v_a = \frac{\nu_a}{\nu_5}, \quad   a=1,2,3, \label{new_couplings}
\ee

In contrast to the $(2+1)$-dimensional model the level of computational complexity in $(3+1)$-dimensional theory is so high that the final result cannot be attained by the usual Feynman diagrammatic technique. Only the renormalization of $G$ and $\lambda$ was possible by directly calculating the diagrams \cite{towards}\footnote{This was done by the renormalization of the $\langle N^iN^j\rangle$ propagator on flat space background, that is by finding the divergences of one-loop diagrams with two external ${\cal N}^i$-legs in contrast to the diagrams with external $H_{ij}$-legs used Sect.\ref{AF2+1}.}, whereas for the renormalization of the rest of coupling constants one should use a more efficient method based on Schwinger--DeWitt
\cite{Schwinger,Sch-DeWitt,DeWitt:2003pm,PhysRep,twoloop,Scholarpedia}
or Gilkey--Seeley \cite{Gilkey-Seeley,Avramidi,Vassilevich} heat kernel
technique and its extension in the form of the method of universal functional traces \cite{PhysRep,twoloop}. The latter, along with a special dimensional reduction transform, turns out to be necessary in order to circumvent the problem of {\em nonminimal} and higher-derivative operators. Application of these methods runs as follows.

For the renormalization of the potential part of the action (\ref{action31}) it is sufficient to consider the metric background on which all its five
tensor structures are nonvanishing and can be distinctly
separated. This is the spacetime metric with a generic {\em static}
3-dimensional part $g_{ij}({\bf x})$, and vanishing shift functions
$N^i=0$. Static nature of $g_{ij}$ and zero shift functions lead to
zero kinetic term of (\ref{action31}) whose contribution is not needed
for the renormalization of couplings $\nu_1,...\nu_5$. The one-loop effective action on such a background is given by Gaussian integration over the full set of quantum fields $(h_{ij},n^i,c^i,\bar c_i)$. These fields in the $(\sigma,\xi)$-family of background covariant gauges of Sect.\ref{BRST}
have the following quadratic parts of their gauge-fixed and ghost actions,
\begin{align}
&S_h = \frac{1}{2G}\int d\tau d^3x \sqrt{g}\,
h_{mn}G^{mn,ij}\left[-\delta_{ij}^{\;\;\;kl}\partial_\tau^2 + {\bf D}_{ij}^{\;\;\;kl}(\nabla)
\right]\,h_{kl}, \label{h_part}\\
&S_n=\frac{\sigma}{2G}\int d\tau d^3x \sqrt{g}\,n^i {\cal O}_{ij}\left[ -
  \delta^j_k\partial_\tau^2 +  {\bf B}^j_{\;\;k}(\nabla)\right] n^k, \label{shift_part}\\
&S_{\rm gh} =
\frac1G \int d\tau d^3x\sqrt{g}\,
\bar{c}_i \left[-\delta^i_j\d_\tau^2 + {\bf B}_{~j}^i (\nabla)\right]c^j.
\end{align}
It is important that for these gauges the $n^ih_{kl}$ cross term is absent in the quadratic part of the total action, and for a static background the spatial differential operator ${\bf B}_{~j}^i (\nabla)$ is exactly the same in the shift and ghost parts of the action. The spatial differential operators here have the form,
\begin{eqnarray}
\label{Dtildeflat}
{\bf D}_{ij}^{\;\;\;kl}(\nabla)&=&-\bigg[\,
\nu_5\delta_{ij}^{\;\;\;kl}
+\frac{4\nu_4(1-\l)+\nu_5}{1-3\l}g_{ij}g^{kl}\bigg]\Delta^3
+\bigg[\,2\nu_5-\frac{1}{\s}\,\bigg]
\delta_{(i}^{(k}\nabla_{j)}\nabla^{l)}\Delta^2\nonumber\\
&&+\frac{4\nu_4(1-\l)+\nu_5}{1-3\l}\,g_{ij}
\nabla^{(k}\nabla^{l)}\Delta^2
+\bigg[4\nu_4+\nu_5+\frac{\l(1+\xi)}{\s}\,\bigg] \nabla_{(i} \nabla_{j)}g^{kl}\Delta^2\nonumber\\
&&-\bigg[\,4\nu_4+2\nu_5+\frac{\xi}{\s}\bigg]\nabla_{(i}\nabla_{j)}
\nabla^{(k} \nabla^{l)}\Delta+\cdots\,,\\
{\bf B}^i_{\;\;j}(\nabla)& =&-\frac1{2\sigma}\delta^i_j\Delta^3
-\frac1{2\sigma}\Delta^2\nabla_j\nabla^i
-\frac{\xi}{2\sigma}\nabla^i\Delta\nabla^k\nabla_j\nabla_k\nonumber\\
&&-\frac{\xi}{2\sigma}\nabla^i\Delta\nabla_j\Delta
 +\frac{\lambda}{\sigma}\Delta^2\nabla^i\nabla_j
 + \frac{\lambda\xi}{\sigma}\nabla^i\Delta^2\nabla_j, \quad
\Delta=\gamma^{ij}\nabla_i\nabla_j,
\end{eqnarray}
where in the metric sector we retain only the principal symbol part of the operator (its highest order derivative terms), because the rest of it contains about two hundred lower derivative terms\footnote{This directly points out to the level of calculational complexity of the problem, which even in the framework of background field and heat kernel approach can be solved only with the aid of {\em xAct} package of {\em Mathematica} program \cite{3+1}.} with coefficients linear, quadratic and cubic in spatial curvature and its covariant derivatives $\nabla_i$ (defined with respect to the background metric $g_{ij}$ --- the notation used in contrast to $D_i$ above). The notation $G^{ij,kl}$ is used for the DeWitt metric (modified by generic $\lambda$) in the space of second rank tensor fields
\be
\label{DeWittG}
G^{ij,kl}=\frac{1}{8}(g^{ik}g^{jl}+g^{il}g^{jk})
-\frac{\l}{4}g^{ij}g^{kl}\;,~~~~~~
\delta_{ij}^{\;\;\;kl}\equiv\delta_{(i}^{k}\delta_{j)}^{l}\;.
\ee

Gaussian integration over $(h_{ij},n^i,c^i,\bar c_i)$ gives the one-loop effective action
\be
\varGamma_{\rm one-loop}=\frac12\,{\rm Tr}\,\ln\left[-\delta_{ij}^{\;\;\;kl}\partial_\tau^2 + {\bf D}_{ij}^{\;\;\;kl}(\nabla)\right]
-{\rm Tr}\,\ln\left[-\delta^i_j\d_\tau^2
+ {\bf B}_{~j}^i (\nabla)      \label{efeq}
\right]\,,
\ee
where the contribution of the gauge-fixing matrix ${\cal O}_{ij}$ is cancelled by the extra normalization factor $({\rm Det}\, {\cal O}_{ij})^{1/2}$ which comes from smearing the gauge-fixing conditions with a Gaussian weight (that generates the gauge-breaking term (\ref{local}), see discussion in Sect.\ref{BRST}) and the contribution of $G^{ij,kl}$ is cancelled by the local measure.

Thus the operators in (\ref{efeq}), usually treated under the sign of functional trace-logarithm by the heat kernel method, are both strongly nonminimal -- their derivatives do not form a Laplacian or d'Alembertian -- and go beyond the second order in derivatives. Therefore they cannot be directly handled by the Schwinger-DeWitt heat kernel method which applies only to minimal second order operators. The first step to circumvent this difficulty is to use the following dimensional reduction transform for the set of operators ${\bf{F}}=({\bf D}_{ij}^{\;\;\;kl},{\bf B}_{~j}^i )$, which is possible for static in time backgrounds,
\begin{align}
\label{Ssqrt}
\frac12 {\rm Tr}\,\ln\big[\!&-\d_\tau^2+
{\bf{F}}\,\big]\nonumber\\
=&-\frac{1}{2}\int d\tau\, d^3x\int_0^\infty \frac{ds}{s}\,{\rm
  tr}\,e^{-s(-\d_\tau^2+
{\bf{F}})}\int_{-\infty}^\infty\frac{d\omega}{2\pi} e^{i\omega(\tau-\tau')}
\delta({\bf x},{\bf x}')\,\Big|_{\,\tau=\tau',\,{\bf x}={\bf x}'}\nonumber\\
=&-\frac{1}{4\sqrt{\pi}}\int d\tau\,d^3x\int_0^\infty \frac{ds}{s^{3/2}}\,
{\rm tr}\,e^{-s{\bf{F}}}
\delta({\bf x},{\bf x}')\,\Big|_{\,{\bf x}={\bf x}'}\nonumber\\
=&-\frac{\Gamma(-1/2)}{4\sqrt{\pi}}\int d\tau\,d^3x\,
{\rm tr}\,\sqrt{{\bf{F}}}\,
\delta({\bf x},{\bf x}')\,\Big|_{\,{\bf x}={\bf x}'}
=\frac{1}{2}\int d\tau\;  {\rm Tr}_3\sqrt{{\bf{F}}}\;.
\end{align}
Here $\rm tr$ denotes the matrix trace in the vector space of field indices and ${\rm Tr}_3$ is the functional trace on the space of functions of 3-dimensional coordinates, as opposed to the four-dimensional ${\rm Tr}\equiv {\rm Tr}_4$.

To handle the operator square root we note that the sixth order operators $\bf{F}$ have the form
 \begin{eqnarray}
{\bf F}(\nabla)=\sum_{a=0}^6 {\cal R}_{(a)}\!
    \sum_{6\geq 2k\geq a}\alpha_{a,k}\nabla_{1}...\nabla_{2k-a}
    (-\Delta)^{3-k},\quad
    {\cal R}_{(a)}=O\Big(\,\frac1{l^a}\,\Big),
\end{eqnarray}
where ${\cal R}_{(a)}$ denote the local coefficients built of curvature tensor and its covariant derivatives of (physical) dimension $a$ in units of inverse length. Obviously, its square root is a pseudodifferential operator which, when expanded in powers of the curvature, becomes an infinite series in local curvature structures ${\cal R}_{(a)}$ of ever growing dimensionality $a$, such that the coefficient of every such structure is an operator polynomial in covariant derivatives of a finite order (determined by $a$) times a certain power of the covariant Laplacian $\Delta$. Dimensional considerations suggest that such an expansion, containing a finite set of positive powers of $\Delta$ and an infinite sequence of its negative powers, looks as follows,
    \begin{eqnarray}
    \sqrt{{\bf F}(\nabla)}
    =\sum\limits_{a=0}^\infty
    {\cal R}_{(a)}\sum\limits_{k\geq a/2}^{K_a}\tilde\alpha_{a,k}
    \nabla_{1}...\nabla_{2k-a}
    \frac1{(-\Delta)^{k-3/2}},    \label{exp_sqrt}
    \end{eqnarray}
where $K_a$ is some finite integer for any $a$ and $\alpha_{a,k}$ are some coefficients.

The sequence of these tensor structures ${\cal R}_{(a)}$ and coefficients $\alpha_{a,k}$ can be found by iterations. First, one writes the needed square root as a sum of the term of zeroth order in curvature ${\bf Q}^{(0)}$  and the perturbation ${\bf X}$, $\sqrt{\bf F}={\bf Q}^{(0)}+{\bf X}$. ${\bf Q}^{(0)}$ is built solely in terms of covariant derivatives and the powers of $\Delta$. Zeroth order ${\bf Q}^{(0)}$ can be chosen by the following procedure. Take the principal symbol ${\cal F}(p)$ of the operator ${\bf F}(\nabla)$ by discarding all its curvature terms and replacing all covariant derivatives $\nabla_i$ by $c$-number (momentunm) vectors $p_i$,
\be
{\cal F}(p)={\bf F}(\nabla)\;
\big|_{\,\nabla\to p,\,{\cal R}\to\,0},
\ee
and extract the matrix square root out of ${\cal F}(p)$. Then define ${\bf Q}^{(0)}$ by replacing back vector arguments $p_i$ by the covariant derivatives $\nabla_i$ with their arbitrary but fixed once and for all ordering,
\be
{\bf Q}^{(0)}(\nabla)=\big[{\cal F}(p)\big]^{1/2}\,\big|_{\,p\to\,\nabla}.
\ee

Then use the fact that the unknown perturbation operator ${\bf X}(\nabla)$ satisfies the equation
\begin{eqnarray*}
{\bf Q}^{(0)}{\bf X}+{\bf X}\,{\bf Q}^{(0)}
    ={\bf F}-\big({\bf Q}^{(0)}\big)^2-
    {\bf X}^2\;.
\end{eqnarray*}
With this choice of ${\bf Q}^{(0)}$ the right hand side here contains a ``source term'' ${\bf F}-({\bf Q}^{(0)})^2\sim[\nabla,\nabla]\sim {\cal R}$ which is at least linear in the curvature, because this difference can be nonzero only due to noncommutativety of covariant derivatives. Discarding the ${\bf X}^2$ term in the right hand side of this equation we see that in the lowest order ${\bf X}$ satisfies the linear equation and has a solution linear in ${\cal R}$. Including ${\bf X}^2$ and solving the equation by iterations then gives a systematic method of expanding the needed square root $\sqrt{\bf F}$ in powers of the curvature and its derivatives. This expansion will have the form (\ref{exp_sqrt}) with all powers of $\Delta$ commuted to the right, which can always be done, because every commutator $[\nabla,\Delta]\propto{\cal R}$ generates extra power of the curvature.

Thus in view of (\ref{Ssqrt}) the divergent part of the effective action (\ref{efeq}) takes the form of the sum of (integrated over time) terms
 \begin{eqnarray}
{\rm Tr}_3\sqrt{\bf F}\,\big|^{\,\rm div}
    =\sum\limits_{a=2}^6\sum\limits_k\,
\tilde\alpha_{a,k}
\int d^3x\,
    {\cal R}_{(a)}({\bf x})
    \nabla_{1}...\nabla_{2k-a}
    \frac1{(-\Delta)^{k-3/2}}\delta({\bf x},{\bf x}')\,\Big|_{\,{\bf
        x}={\bf x}'}^{\,\rm div},
\end{eqnarray}
which are just the universal functional traces of \cite{PhysRep}. Calculation of these UV divergences is based on the proper time representation and the heat kernel of the minimal second order operator -- the covariant Laplacian $\Delta$,
 \begin{eqnarray}
\nabla...\nabla\frac{\hat 1}{(-\Delta)^\alpha}\,\delta({\bf x},{\bf x}')\,
\Big|_{\,{\bf x}'={\bf x}}^{\,\rm div} =\frac{1}{\varGamma(\alpha)}
\nabla...\nabla\int_0^\infty ds\,s^{\alpha-1}\,e^{s\Delta}\,
\hat\delta({\bf x},{\bf x}')\,
    \Big|_{\,{\bf x}'={\bf x}}^{\,\rm div}\,.
\end{eqnarray}
The expansion at $s\to 0$ of the heat kernel $e^{s\Delta}\,\hat\delta({\bf x},{\bf x}')$ in terms of the Synge world function $\sigma({\bf x},{\bf x}')$ and HAMIDEW coefficients $a_n({\bf x},{\bf x}')$ allows one to systematically isolate UV divergences of these quantities as the divergence of the proper time integral $\int_0^\infty ds/s$, due to well-known coincidence limits of $\sigma({\bf x},{\bf x}')$, $a_n({\bf x},{\bf x}')$ and their multiple derivatives \cite{Sch-DeWitt,PhysRep}.

With the identification
\begin{align}
\int_0^\infty\frac{\di s}s\mapsto\log\left(\frac{\Lambda_{\rm UV}^2}{k_*^2}\right)\;,
 \end{align}
(it differs from (\ref{cutoff2}) because the dimension of $s$ now is $-2$) this leads to the same Eqs.(\ref{Gnu_0})-(\ref{beta_Gnu}) for renormalized couplings and their beta-functions with very complicated set of functions $C_G$ and $C_{\nu_a}$ in $(3+1)$-dimensional model. The resulting $\beta_\lambda$ (obtained in \cite{towards} by the diagrammatic method) reads
\be
\beta_\lambda=\frac{\cal G}{120\pi^2}\,
    \frac{27(1-\lambda)^2
    +3u_s(11-3\lambda)(1-\lambda)
    -2u_s^2(1-3\lambda)^2}{(1-\lambda)(1+u_s)u_s}\,, \label{beta_l}
\ee
while the rest of beta-functions of essential couplings ${\cal G}$ and $\chi = (u_s,v_1,v_2,v_3)$ are given by the expressions
\begin{eqnarray}
&&\beta_{\cal G}  =\frac{{\cal G}^2}{26880\pi^2} \frac{ \sum_{n=0}^7 u_s^n\, {\cal P}^{\cal
    G}_n[\l,v_1,v_2,v_3]}{(1-\lambda)^2(1-3\lambda)^2
    (1+u_s)^3 u_s^3} ,                    \label{beta_calG}\\
    &&\beta_\chi  =\frac{{\cal G}}{26880\pi^2} \frac{A_\chi \sum_{n=0}^9 u_s^n\,
    {\cal P}^{\chi}_n[\l,v_1,v_2,v_3]}
    {(1-\lambda)^3(1-3\lambda)^3
    (1+u_s)^3 u_s^5} ,                   \label{beta_chi}
\end{eqnarray}
where $A_{u_s} = u_s (1-\lambda)$, $A_{v_1} = 1$, $A_{v_2}=A_{v_3} = 2$ and ${\cal P}^{{\cal G},\chi}_n[\l,v_1,v_2,v_3]$ is a set of very complicated polynomials of its arguments, which takes pages \cite{3+1}. For illustration, an example of such a polynomial (one of the longest ones) is
\begin{eqnarray}
{\cal P}^{v_1}_5& =&-2(1 - \lambda)^2(1-3\lambda) \Bigl\{ 168 v_2^3 (51 \lambda ^3-149 \lambda ^2+125 \lambda -27 )  -108 v_3^3 (9 \lambda ^3+9 \lambda^2\notag\\
&&-25 \lambda +7 )   -4 v_2^2  (1-\lambda )   \bigl[ 18 v_3 (117 \lambda ^2-366 \lambda +109 )   -284 \lambda ^2-7265 \lambda+5425 \bigr]\notag\\
&&+40320 v_1^2 (1-\lambda )^2 (\lambda +1) -9 v_3^2 (3467 \lambda ^3-8839 \lambda ^2+6237 \lambda -865 ) \notag\\
&&+v_1  \bigl[ 64  v_2^2(1-\lambda)^2 (1717 \lambda -581) - 16 v_2 (1-\lambda ) \bigl( 3 v_3  (2741 \lambda ^2-3690 \lambda +949 )\notag\\
&&+25940 \lambda ^2-40662 \lambda+12022 \bigr) +27 v_3^2 (961 \lambda ^3-2395 \lambda ^2+1835 \lambda -401 )\notag\\
&&+6  v_3 (52267 \lambda ^3-148963 \lambda ^2+129881 \lambda -33185 ) -288353 \lambda ^3+542255 \lambda ^2\notag\\
&&-333355 \lambda+83485 \bigr] -2  v_2  \bigl[ 162 v_3^2 (3 \lambda ^3+35 \lambda ^2-51 \lambda +13 )  + 24  v_3  (1265 \lambda ^3\notag\\
&&-2191 \lambda ^2+691 \lambda +235 ) +30971 \lambda ^3-40323 \lambda ^2+13167 \lambda-4451 \bigr]\notag\\
&&-12  v_3  (6551 \lambda ^3-11593 \lambda ^2+6124 \lambda -1112) +109519 \lambda^3-252396 \lambda ^2\notag\\
&&+177357 \lambda -34396\,\Bigr\}\,.
\end{eqnarray}

\subsection{Dimensional regularization in Ho\v{r}ava models}
It is instructive to compare the Wilsonian type beta-functions (\ref{beta_Gnu}), obtained by differentiation with respect to the running scale $k_*$, and beta functions in minimal subtraction scheme of the dimensional regularization \cite{t`Hooft,WeinbergQFT}, especially bearing in mind that the dimensions of coupling constants are defined now with respect to Lifshitz dimensionality rather than the physical one.

For generic multicharge theory with the {\em dimensionless renormalized} coupling constants $g=\{\,g^A\,\}$, $A=1,2,...N$, their bare couplings are denoted as $g_0^A$ and have nontrivial dimensions which get regularized according to
    \begin{eqnarray}
    &&[\,g_0^A]=\Delta_A+\rho_A\varepsilon, \quad [\,g^A]=0, \quad
    \varepsilon=d_{\rm phys}-d\to 0,      \label{dims}
    \end{eqnarray}
where $d_{\rm phys}$ is a physical (integer) space dimensionality and $d$ is is its regularization by analytical continuation into the complex plane. These dimensions are generically linear in $\varepsilon$ with specific coefficients $\rho_a$ \cite{WeinbergQFT}. The bare coupling is a series in powers of $1/\varepsilon$ as a function of renormalized couplings
    \begin{eqnarray}
    g_0^A=\mu^{\Delta_A+\rho_A\varepsilon}\left(g^A
    +\sum\limits_{n=1}^\infty
    \frac{a_n^A(g)}{\varepsilon^n}\right), \label{poles}
    \end{eqnarray}
where $\mu$ is a running scale whose factor recovers a correct dimension of the bare constant. One loop approximation contributes a single pole in $\varepsilon$, while higher order poles are due to multi-loop corrections.

A conventional assumption is that $g_0$ is independent of $\mu$, $(\mu d/d\mu)g_0^a=0$, and divergent at $\varepsilon\to 0$ while the running renormalized charge $g=g(\mu,\varepsilon)$ is {\em analytic at $\varepsilon\to 0$}. This leads to the definition of the beta function which is also analytic in this limit, $(\mu d/d\mu)g^A(\mu,\varepsilon)= \beta^A(g(\mu,\varepsilon),\varepsilon)$. Then, differentiating (\ref{poles}) and demanding the cancellation of the linear in $\varepsilon$, $\varepsilon^0$ and $\varepsilon^{-1}$ terms, one has \cite{t`Hooft,WeinbergQFT}
    \begin{eqnarray}
    \beta^A(g,\varepsilon)=-(\Delta_A+\rho_A\,\varepsilon)\,g^A-\rho_A\, a_1^A(g)+\sum\limits_{B=1}^N\rho_B\, g^B
    \frac{\partial a_1^A(g)}{\partial g^B},     \label{dimbeta}
    \end{eqnarray}
whereas the cancellation of higher order poles results in recurrent equations for $a_n$, $n>1$.

In context of Ho\v{r}ava gravity $g^A\mapsto(G,\nu^a)$, and $\lambda$ can be included into the set of $\nu_a$ just like it was done in Eq.(\ref{Gnu_0}) for $(2+1)$-dimensional model. The dimension parameters of Eqs.(\ref{dims})-(\ref{poles}) should be interpreted in terms of anisotropic Lifshitz dimensions. In particular, the Lifshitz dimension of coupling constants should be related to the regularized space dimensionality
    \be
    [\,G_0\,]=d_{\rm phys}-d=\varepsilon,\quad [\,\nu^0_a\,]=0.
    \ee
Note that the coupling constant $G$ differs from the rest of the couplings because the dimension of its bare version gets regularized. In accordance with the equation (\ref{dims}) this identifies its parameters to be $\rho_G=1$, $\rho_a=0$ and $\Delta_G=\Delta_a=0$.

Within minimal subtraction scheme the pole parameter $\varepsilon$ is related to the UV cutoff of Eq.(\ref{Gnu_0}), $\ln(\varLambda^2_{\rm UV}/k_*^2)=2/\varepsilon$, so that this equation can be rewritten in the form of Eq.(\ref{poles})
    \be
    G_0=\mu^{\varepsilon}\left(G
    +\frac{4\,C_G\,G^2}{\varepsilon}+...\right),\quad
    \nu_a^0=\nu_a+\frac{4\,G\,(C_G\nu_a-C_{\nu_a})}\varepsilon+...,
    \ee
which means that $a_1^G=4C_GG^2$ and $a_1^{\nu_a}=4G(C_G\nu_a-C_{\nu_a})$. Therefore, according to (\ref{dimbeta}) and in view of the fact that $a_1^G$ and $a_1^{\nu_a}$ are respectively quadratic and linear in $G$ (remember that one-loop $C_G$ and $C_{\nu_a}$ are $G$-independent) one has
\begin{align}
&\beta_G=\bigg[-\varepsilon\,G-a_1^G+G\frac{\partial a_1^G}{\partial G}\bigg]_{\,\varepsilon=0}=a_1^G=
4\,C_G\,G^2,\\
&\beta_{\nu_a}=G\frac{\partial a_1^{\nu_a}}{\partial G}=a_1^{\nu_a}
= -4 G C_{\nu_a}
+\nu_a\frac{\beta_G}G,
\end{align}
which fully agrees with the Wilsonian beta-functions (\ref{beta_Gnu}). Both beta functions again (as in Wilsonian approach) coincide with single pole residues of the bare constants in terms of renormalized ones, but this is achieved via nontrivial combination of terms in (\ref{dimbeta}) and their special scaling in the perturbation theory constant $G$.

\subsection{Fixed points}
The number of coupling constants and complexity of their beta-functions thus far preclude from the full analysis of RG flows in $(3+1)$-dimensional Ho\v{r}ava model. However, preliminary observations of their properties allow one to come to interesting conclusions and further prospects of this model.

An important question is the existence and nature of fixed points of
the RG flow. The dependence of the $\beta$-functions (\ref{beta_l})-(\ref{beta_chi}) on the coupling ${\cal G}$ factorizes, and this coupling determines the overall strength of interactions and must be
small for the validity of the perturbative expansion. Its UV behavior
determines whether the model is asymptotically free (${\cal G}\to 0$)
or has a Landau pole (${\cal G}\to\infty$). The rest of the couplings $\l,u_s, v_a$ are ratios of the coefficients in the action and need not be small. The search for fixed points of the RG flow thus consists in finding them for a subspace of the couplings $\l,u_s, v_a$ by solving the system,
\be
\beta_\lambda/{\cal G}=0\;,\quad
\beta_\chi/{\cal G}=0\;,~~~~~ \chi=u_s,v_1,v_2,v_3, \label{fpeq}
\ee
and then evaluating $\beta_{\cal G}$ at a given solution, whose sign determines whether the flow trajectory goes to a Gaussian fixed point or a Landau pole. The results are summarized in the Table \ref{tabFP1}. All these
fixed points turn out to be asymptotically free, but the two last points correspond to very large values of $v_1$ and their validity should be taken with a certain reservation.
\begin{table}[h]
\begin{center}
 \begin{tabular}{| c | c | c | c | c | c |c |c|}
 \hline
$\lambda$  &$u_s$ & $v_1$ & $v_2$ & $v_3$ &  $\beta_{\cal G}/{\cal
  G}^2$ & AF?
 &UV attractive along $\lambda$?\\ [0.5ex]
 \hline\hline
0.1787&  60.57 &-928.4 & -6.206 & -1.711 &  -0.1416 & yes& no \\ [0.5ex]
  \cline{1-8}
 0.2773 & 390.6  &-19.88& -12.45 & 2.341 &  -0.2180  & yes & no\\ [0.5ex]
  \cline{1-8}
0.3288 & 54533 & 3.798$\times 10^8$ & -48.66 & 4.736 &  -0.8484 & yes&
no\\ [0.5ex]
  \cline{1-8}
 0.3289 & 57317 &-4.125$\times 10^8$ & -49.17 & 4.734 & -0.8784 &yes &
 no\\ [0.5ex]
 \hline
  \end{tabular}
  \vspace{0.5cm}
  \caption{\label{tabFP1}
Solutions of the system (\ref{fpeq}).
The sixth column gives the value of the
    $\beta$-function for ${\cal G}$ at the respective solution and the
    seventh column indicates whether it corresponds to an
    asymptotically free fixed point. The eighth column tells if the
    fixed point is UV attractive along the $\lambda$-direction.}
\end{center}
\end{table}
\vspace{-0.5cm}

It was conjectured in \cite{Gumrukcuoglu:2011xg} that the UV fixed
points of HG can be at infinite $\l$ and that the limit $\l\to\infty$
is interesting in cosmological applications. This turns out to be true ---  all $\beta$-functions are finite at $\l\to\infty$, whereas $\beta_\l$ is proportional to $\l$, $\beta_\l=-3\l\,{\cal G}(3-2u_s)/40\pi^2 u_s$, $\l\to\infty$. Fixed points at $\l=\infty$, which solve the equations $\beta_\chi/{\cal G}\,|_{\,\l=\infty}=0$ for $\chi=u_s,v_1,v_2,v_3$, are collected in Table~\ref{tabFP2}. Three among them are UV attractive along the $\l$-direction and correspond to asymptotically free fixed points. The structure of the RG flow around these points deserves a detailed study, which is also very interesting in the context of the Perelman--Ricci flows \cite{Frenkel:2020dic}.

\begin{table}[h]
\begin{center}
\begin{tabular}{| c | c | c | c | c | c |c |}
 \hline
$u_s$  & $v_1$ & $v_2$ & $v_3$ &   $\beta_{\cal G}/{\cal G}^2$&  AF?
 &UV attractive along $\l$?\\ [0.5ex]
\hline\hline
0.01950& 0.4994 & -2.498 & 2.999 &  -0.2004 &yes&no\\ [0.5ex]
\cline{1-7}
 0.04180 & -0.01237 & -0.4204 & 1.321 &  -1.144&yes&no \\ [0.5ex]
  \cline{1-7}
0.05530 & -0.2266 & 0.4136 & 0.7177 &  -1.079 &yes&no\\ [0.5ex]
 \cline{1-7}
12.28 & -215.1 & -6.007 & -2.210 &   -0.1267 &yes&yes\\ [0.5ex]
 \cline{1-7}
21.60 & -17.22 & -11.43 & 1.855 &   -0.1936 &yes&yes\\ [0.5ex]
 \cline{1-7}
440.4 & -13566 & -2.467 & 2.967 &   0.05822 &no&yes\\ [0.5ex]
 \cline{1-7}
571.9 & -9.401 & 13.50 & -18.25 &  -0.07454 &yes&yes\\ [0.5ex]
 \cline{1-7}
950.6 & -61.35 & 11.86 & 3.064 &   0.4237 &no&yes\\ [0.5ex]
 \hline
  \end{tabular}
  \vspace{0.5cm}
  \caption{\label{tabFP2} Fixed points of Ho\v rava gravity at $\l=\infty$. }
\end{center}
\end{table}
Another interesting observation is that for a special choice of the values
\be
\{v^*\}: v_1 = 1/2,\quad v_2 = -5/2, \quad v_3 = 3,
 \label{crit_point}
\ee
the limit $u_s\to 0$ of all $\beta$-functions except $\beta_\l$
becomes regular for any $\l$ in the unitary domain and, moreover, three beta-functions turn out to be zero, $\beta_{v_a}\big|_{\{v^*\},\,u_s\to 0}=0$, $a=1,2,3$. The point $\{v^*\},\;u_s\to 0$ is special since it corresponds to the version of HG, in which the potential term is a square of the Cotton tensor $C_{ij}=\varepsilon^{kl(i}\nabla_k R^{j)}_{\,l}$, $\varepsilon^{ikl}=\epsilon^{ikl}/\sqrt{g}$, $\epsilon^{123}=1$.
\begin{align}
S[\,g\,]= \frac{1}{2G} \int \di\tau\, \di^3x \,
\sqrt{g}\,(K_{ij}K^{ij}
-\lambda K^2 + \nu_5\,C^{ij}C_{ij}). \label{db}
\end{align}
This version of HG was originally suggested in \cite{Horava:2009uw}
and its quantum properties were studied in \cite{Orlando:2009en}. It is
known as HG with detailed balance and is interesting because
the Cotton tensor can be rewritten as a variational derivative of the
3-dimensional gravitational Chern--Simons action, $\sqrt{g}\,C^{ij}=-\delta W_{\rm CS}[\,g\,]/\delta g_{ij}(x)$,
\begin{eqnarray}
&&W_{\rm CS}[\,g\,]=\frac{1}{2}\int
\di^3x\,\epsilon^{ijk}\Big(\,
\Gamma^m_{il}\partial_j\Gamma^l_{km}+
\frac23\Gamma^n_{il}\Gamma^l_{jm}\Gamma^m_{kn}\Big),  \label{actionCS}
\end{eqnarray}
defined in terms of the Christoffel symbol as a functional of
$g_{ij}$. Then the integrand of (\ref{db}) can be rewritten as a square of the Langevin equation characteristic of stochastic quantization of 3-dimensional gravity \cite{Parisi:1980ys,Damgaard:1987rr}. Further, there exists a deformation of the action
(\ref{actionCS}) by relevant operators which preserves the detailed balance
structure and is related to the topological massive gravity
\cite{Deser:1982vy,Deser:1981wh,Deser:1990bj}. The detailed balance
relation between $d$ and $(d+1)$-dimensional theories appears in the
context of stochastic quantization and establishes a nontrivial
connection between the renormalization properties of the two theories
\cite{Zinn-Justin:1986nph}. In our case this suggests an intriguing
connection between the $(3+1)$-dimensional projectable HG and the
$3$-dimensional gravitational Chern--Simons / topological massive
gravity \cite{Orlando:2009en}.

It is important to emphasize, however, that the point $\{v^*\},\,u_s\to
0$ is neither fixed nor fully regular point of the RG flow, because
the $\beta$-function of the remaining essential coupling $\l$ diverges in this limit, $\beta_{\l}\,|_{\{v^*\},\,u_s\to 0}\sim 1/u_s$.
Thus, the physical significance of the critical point (\ref{crit_point}) is
unclear at the moment. It will be interesting to understand if the
inclusion of fermionic degrees of freedom appearing in the stochastic
quantization framework \cite{Orlando:2009en} can change the picture.

\section{Conclusions and discussion}
In this overview we have briefly exposed the current status of UV renormalization in Ho\v{r}ava gravity theory. This includes the renormalizability of its projectable models in all spacetime dimensions, subtle mechanism of a possible restoration of unitarity in their non-projectable version, demonstrates the origin of UV asymptotic freedom in $(2+1)$-dimensional HG and points out to the existence of several fixed points of RG flow in $(3+1)$-dimensional theory, which also might serve as good candidates for its UV asymptotic freedom. Along with the description of the calculational strategy for beta-functions of the latter theory we dwelt on the generalized Schwinger-DeWitt technique of universal functional traces which strongly exceeds conventional Feynman diagrammatic method in its capacity to handle the models with broken Lorentz symmetry. The combination of this method with the heat kernel approach and background field formalism makes tractable computations in theories encumbered with a plethora of spacetime anisotropic structures inherent in HG.

In addition to implications of HG theory listed in Introduction it is also worth mentioning other studies spanning several directions. As it has already been mentioned, in \cite{Orlando:2009en} the projectable version in $d=3$ was considered with the detailed balance restriction on the parameters of the model \cite{Horava:2009uw}. This model is connected to 3-dimensional topologically massive gravity via the stochastic quantization approach and it is argued that it inherits the renormalizability properties of the latter --- the conclusion matching with the properties of beta-functions and their fixed points discussed in Sect.\ref{3+1}. However, the treatment of HG gauge invariance in \cite{Orlando:2009en} is somewhat obscure. The works \cite{Horava:2009if,Anderson:2011bj,Ambjorn:2010hu,Sotiriou:2011mu,Benedetti:2014dra}
study the relation between HG and causal dynamical triangulations. In Ref.~\cite{Benedetti:2013pya} a one-loop renormalization of a truncated version of the $d=2$ projectable model was considered, but this type of truncation explicitly breaks gauge invariance of the theory. Finally, in Refs.~\cite{D'Odorico:2014iha} the one-loop counterterms for the
gravitational effective action induced by a scalar field with
Lifshitz scaling (see also \cite{Giribet:2010th,Nesterov:2010yi,Baggio:2011ha} for earlier works on this subject) were computed. These counterterms were shown to have the
structure of the bare HG action, which suggests that if pure HG is renormalizable, it remains so upon inclusion of matter. In fact this property follows from gauge-fixing procedure of Sec.\ref{reg_prop} --- the choice of gauge for Lifshitz matter gauge symmetry preserving regularity of propagators is also possible \cite{HG}.

Besides applications in quantum gravity, Ho\v rava gravity in $d=2$ can govern the dynamics of membranes in M-theory \cite{Horava:2008ih}. Other
applications include the holographic description of non-relativistic
strongly coupled systems, analogous to those occurring in condensed matter
physics \cite{Janiszewski:2012nb,Griffin:2012qx}, the model of multilayer graphene \cite{Volovik}.

To finish this review of quantum Ho\v{r}ava gravity it is perhaps worth adding some comments relating its status to string theory.
String theory provides a fruitful approach to the construction of a consistent theory of quantum gravity, see e.g. \cite{Polchinski}. But it makes this at the expense of introducing a very rich extra structure, complexity and widely recognized nonuniqueness \cite{landscape}. As opposed to this nonuniqueness and richness of string theory, overloaded with numerous extra structures, it makes sense to question if quantum gravity can be self-contained and consistent in a smaller framework. HG is an attempt to formulate such a framework constituted by the requirements of locality, renormalizability and unitarity in $(3+1)$ dimensions. Ho\v{r}ava gravity seems to be the only known now example satisfying this set of requirements, and if it would be asymptotically free, that is consistently complete in UV domain, and extended to the realm of non-projectable models interpolating between low energy GR and Planckian physics, then it will have good chances of being the physical theory of our Nature. The obtained results is a strong indication to the plausibility of this conjecture.

\section*{Acknowledgement}
Original methods exhibited in this review would not be possible without strong thought provoking influence of B. S. DeWitt and G. A. Vilkovisky to whom I am deeply grateful. I am also deeply grateful to D. Blas, M. Herrero-Valea, A. V. Kurov, S. M. Sibiryakov and C. F. Steinwachs for long term collaboration on the original results of this paper, especially emphasizing Sergey Sibiryakov as a moving spirit behind the studies of Lorentz symmetry violating gravitational models. I would like to thank J. Bellor\'{\i}n, I. L. Buchbinder, J. Donoghue, M. Duff, G. Dvali, S. Fulling, A. Yu. Kamenshchik, E. Mottola, V. Mukhanov, S. Mukohyama, D. V. Nesterov, H. Osborn, N. Ohta, V. A. Rubakov, M. Sasaki, I. L. Shapiro, M. Shaposhnikov, S.~Solodukhin, P. Stamp,  A. A. Starobinsky, K. Stelle, A. A. Tseytlin, M. A. Vasiliev, W. Unruh, W. Wachowski and R. Woodard for fruitful discussions.
%
%
%



\end{document}